\documentclass[twocolumn]{aastex631}

\usepackage{amsmath}

\definecolor{fuchsia}{rgb}{0.54, 0.17, 0.89}
\definecolor{azure}{rgb}{0.0, 0.5, 1.0}
\definecolor{pgreen}{rgb}{0.12, 0.3, 0.17}
\definecolor{alizarin}{rgb}{0.82, 0.1, 0.26}

\newcommand{\kms}{{\rm km~s^{-1}}}

\newcommand{\oiii}{[\textrm{O}~\textsc{iii}]}

\newcommand{\simgt}{\,\rlap{\lower 3.5 pt \hbox{$\mathchar \sim$}} \raise
1pt \hbox {$>$}\,}
\newcommand{\simlt}{\,\rlap{\lower 3.5 pt \hbox{$\mathchar \sim$}} \raise
1pt \hbox {$<$}\,}
\newcommand{\Msun}{M_{\odot}}

\newcommand{\logm}{\log M_*/\Msun}

\newcommand{\hb}{${\rm H\beta}$}

\newcommand{\hst}{{HST}}
\newcommand{\jwst}{{JWST}}

\newcommand{\Z}{{\rm [Z/H]}}

\newcommand{\nsrc}{129} 
\newcommand{\ny}{118} 
\newcommand{\nj}{11} 
\newcommand{\nfld}{19} 
\newcommand{\area}{180} 
\newcommand{\sext}{{\tt SExtractor}}

\submitjournal{ApJ}

\shorttitle{BEACON. I. Survey Overview and Initial Results}
\shortauthors{Morishita, Mason, et al.}

\graphicspath{{./}{figures/}}

\begin{document}

\title{BEACON: JWST NIRCam Pure-parallel Imaging Survey. I. Survey Design and Initial Results}

\correspondingauthor{Takahiro Morishita, Charlotte A. Mason}
\email{takahiro@ipac.caltech.edu, charlotte.mason@nbi.ku.dk}

\author[0000-0002-8512-1404]{Takahiro Morishita}
\altaffiliation{These authors equally contributed to this work.} 
\affiliation{IPAC, California Institute of Technology, MC 314-6, 1200 E. California Boulevard, Pasadena, CA 91125, USA}

\author[0000-0002-3407-1785]{Charlotte A. Mason}
\altaffiliation{These authors equally contributed to this work.} 
\affiliation{Cosmic Dawn Center (DAWN)}
\affiliation{Niels Bohr Institute, University of Copenhagen, Jagtvej 128, DK-2200 Copenhagen N, Denmark}

\author[0009-0005-9953-433X]{Kimi C. Kreilgaard}
\affiliation{Cosmic Dawn Center (DAWN)}
\affiliation{Niels Bohr Institute, University of Copenhagen, Jagtvej 128, DK-2200 Copenhagen N, Denmark}

 \author[0000-0001-9391-305X]{Michele Trenti}
 \affiliation{School of Physics, University of Melbourne, Parkville 3010, VIC, Australia}
 \affiliation{ARC Centre of Excellence for All Sky Astrophysics in 3 Dimensions (ASTRO 3D), Australia}

\author[0000-0002-8460-0390]{Tommaso Treu}
\affiliation{Department of Physics and Astronomy, University of California, Los Angeles, 430 Portola Plaza, Los Angeles, CA 90095, USA}

\author[0000-0003-0980-1499]{Benedetta Vulcani}
\affiliation{INAF Osservatorio Astronomico di Padova, vicolo dell'Osservatorio 5, 35122 Padova, Italy}

\author[0000-0003-3817-8739]{Yechi Zhang}
\affiliation{IPAC, California Institute of Technology, MC 314-6, 1200 E. California Boulevard, Pasadena, CA 91125, USA}

\author[0000-0002-5258-8761]{Abdurro'uf}
\affiliation{Department of Physics and Astronomy, The Johns Hopkins University, 3400 N Charles St. Baltimore, MD 21218, USA}
\affiliation{Space Telescope Science Institute (STScI), 3700 San Martin Drive, Baltimore, MD 21218, USA}

\author[0000-0002-8630-6435]{Anahita Alavi}
\affiliation{IPAC, Mail Code 314-6, California Institute of Technology, 1200 E. California Blvd., Pasadena CA, 91125, USA}

\author[0000-0002-7570-0824]{Hakim Atek}
\affiliation{Institut d'Astrophysique de Paris, CNRS, Sorbonne Universit\'e, 98bis Boulevard Arago, 75014, Paris, France}

\author[0000-0002-3196-5126]{Yannick Bah\'{e}}
\affiliation{Institute of Physics, Laboratory of Astrophysics, Ecole Polytechnique F\'ed\'erale de Lausanne (EPFL), Observatoire de Sauverny, CH-1290 Versoix, Switzerland}

\author[0000-0001-5984-0395]{Maru\v{s}a Brada{\v c}}
\affiliation{University of Ljubljana, Department of Mathematics and Physics, Jadranska ulica 19, SI-1000 Ljubljana, Slovenia}
\affiliation{Department of Physics and Astronomy, University of California Davis, 1 Shields Avenue, Davis, CA 95616, USA}

\author[0000-0002-7908-9284]{Larry D. Bradley}
\affiliation{Space Telescope Science Institute (STScI), 3700 San Martin Drive, Baltimore, MD 21218, USA}

\author[0000-0002-8651-9879]{Andrew J.\ Bunker}
\affiliation{Department of Physics, University of Oxford, Denys Wilkinson Building, Keble Road, Oxford OX1 3RH, UK}

\author[0000-0001-7410-7669]{Dan Coe}
\affiliation{Space Telescope Science Institute (STScI), 3700 San Martin Drive, Baltimore, MD 21218, USA}
\affiliation{Association of Universities for Research in Astronomy (AURA), Inc.~for the European Space Agency (ESA)}
\affiliation{Department of Physics and Astronomy, The Johns Hopkins University, 3400 N Charles St. Baltimore, MD 21218, USA}

\author[0000-0001-6482-3020]{James Colbert}
\affiliation{IPAC, California Institute of Technology, 1200 E. California Blvd, Pasadena, CA 91125, USA}

\author[0000-0001-5487-0392]{Viola Gelli}
\affiliation{Cosmic Dawn Center (DAWN)}
\affiliation{Niels Bohr Institute, University of Copenhagen, Jagtvej 128, DK-2200 Copenhagen N, Denmark}

\author[0000-0001-8587-218X]{Matthew J. Hayes}
\affiliation{Stockholm University, Department of Astronomy and Oskar Klein Centre for Cosmoparticle Physics, AlbaNova University Centre, SE-10691, Stockholm, Sweden.}

\author[0000-0001-5860-3419]{Tucker Jones}
\affiliation{University of California, Davis, Department of Physics and Astronomy, One Shields Avenue, Davis, CA 95616, USA}

\author[0000-0002-2993-1576]{Tadayuki Kodama}
\affiliation{Astronomical Institute, Tohoku University, 6-3 Aramaki, Aoba-ku, Sendai 980-8578, Japan}

\author[0000-0003-4570-3159]{Nicha Leethochawalit}
\affiliation{National Astronomical Research Institute of Thailand (NARIT), Mae Rim, Chiang Mai, 50180, Thailand}

\author[0009-0002-8965-1303]{Zhaoran Liu}
\affiliation{Astronomical Institute, Tohoku University, 6-3 Aramaki, Aoba-ku, Sendai 980-8578, Japan}

\author[0000-0001-6919-1237]{Matthew A. Malkan}
\affiliation{University of California, Los Angeles, Department of Physics and Astronomy, 430 Portola Plaza, Los Angeles, CA 90095, USA}

\author[0000-0001-7166-6035]{Vihang Mehta}
\affiliation{IPAC, California Institute of Technology, 1200 E. California Blvd, Pasadena, CA 91125, USA}

\author[0000-0002-8632-6049]{Benjamin Metha}
\affiliation{School of Physics, University of Melbourne, Parkville 3010, VIC, Australia}
\affiliation{ARC Centre of Excellence for All Sky Astrophysics in 3 Dimensions (ASTRO 3D), Australia}

\author[0000-0001-7769-8660]{Andrew B. Newman}
\affiliation{Observatories of the Carnegie Institution for Science, Pasadena, CA, USA}

\author[0000-0002-9946-4731]{Marc Rafelski}
\affiliation{Space Telescope Science Institute, 
3700 San Martin Drive, 
Baltimore, MD, 21218 USA}
\affiliation{Department of Physics and Astronomy, Johns Hopkins University, Baltimore, MD 21218,USA}

\author[0000-0002-4140-1367]{Guido Roberts-Borsani}
\affiliation{Department of Astronomy, University of Geneva, Chemin Pegasi 51, 1290 Versoix, Switzerland}

\author[0000-0001-7016-5220]{Michael J. Rutkowski}
\affiliation{Minnesota State University, Mankato, Department of Physics and Astronomy, 141 Trafton Science Center N, Mankato, MN 56001, USA}

\author[0000-0002-9136-8876]{Claudia Scarlata}
\affiliation{University of Minnesota, Twin Cities, 116 Church St SE, Minneapolis, MN 55455, USA}

\author[0000-0001-9935-6047]{Massimo Stiavelli}
\affiliation{Space Telescope Science Institute, 3700 San Martin Drive, Baltimore, MD 21218, USA}

\author[0009-0005-1487-7772]{Ryo A. Sutanto}
\affiliation{Astronomical Institute, Tohoku University, 6-3 Aramaki, Aoba-ku, Sendai 980-8578, Japan}

\author[0009-0009-8116-0316]{Kosuke Takahashi}
\affiliation{Astronomical Institute, Tohoku University, 6-3 Aramaki, Aoba-ku, Sendai 980-8578, Japan}

\author[0000-0002-7064-5424]{Harry I. Teplitz}
\affiliation{IPAC, Mail Code 314-6, California Institute of Technology, 1200 E. California Blvd., Pasadena CA, 91125, USA}

\author[0000-0002-9373-3865]{Xin Wang}
\affil{School of Astronomy and Space Science, University of Chinese Academy of Sciences (UCAS), Beijing 100049, China}
\affil{National Astronomical Observatories, Chinese Academy of Sciences, Beijing 100101, China}
\affil{Institute for Frontiers in Astronomy and Astrophysics, Beijing Normal University,  Beijing 102206, China}




\begin{abstract}
We introduce the Bias-free Extragalactic Analysis for Cosmic Origins with NIRCam (BEACON) survey, a JWST Cycle~2 program allocated up to 600 {\it pure-parallel} hours of observations. BEACON explores high-latitude areas of the sky with JWST/NIRCam over $\sim100$\,independent sightlines, totaling $\sim0.3$\,deg$^2$, reaching a median F444W depth of $\approx28.2$\, AB mag (5$\sigma$). Based on existing JWST observations in legacy fields, we estimate that BEACON will photometrically identify 25--150 galaxies at $z>10$ and 500--1000 at $z\sim7$--10 uniquely enabled by an efficient multiple filter configuration spanning $0.9$--5.0\,$\mu$m. The expected sample size of $z>10$ galaxies will allow us to obtain robust number density estimates and to discriminate between different models of early star formation. 
In this paper, we present an overview of the survey design and initial results using the first \nfld\, fields. We present \nsrc\ galaxy candidates at $z\simgt7$ identified in those fields, including \nj\ galaxies at $z\simgt10$ and several UV-luminous ($M_{\rm UV}<-21$\,mag) galaxies at $z\sim8$. The number densities of $z<13$ galaxies inferred from the initial fields are overall consistent with those in the literature. Despite reaching a considerably large volume ($\sim10^5$\,Mpc$^3$), however, we find no galaxy candidates at $z>13$, providing us with a complimentary insight into early galaxy evolution with minimal cosmic variance.
We publish imaging and catalog data products for these initial fields. Upon survey completion, all BEACON data will be coherently processed and distributed to the community along with catalogs for redshift and other physical quantities. 
\end{abstract}



\section{Introduction} \label{sec:intro}

Space telescopes have been essential for understanding galaxy formation and evolution, with the Hubble Space Telescope (\hst) in particular progressively extending our frontiers from the pioneering Hubble Deep Field at redshift $z\sim4$ \citep[lookback time $\sim$12\,Gyr; e.g.,][]{madau96}, to more recent discoveries of galaxies in the epoch of reionization (EoR, \citealp[$z\sim6$--$10$, lookback time $>13$\,Gyr; e.g.,][]{bunker04,oesch10,bouwens10,bunker10,ellis13,bouwens15,oesch18,ishigaki18}). 
However, even with the deepest \hst\ and Spitzer data, key questions remained unanswered: When and how did the first sources of light form? How was intergalactic hydrogen reionized? How are galaxy mass and structure acquired across cosmic time? The James Webb Space Telescope (\jwst) was designed to address these fundamental questions. 

JWST Cycle 1 was a watershed moment. It extended the frontier of galaxy detection to the first 400\,million years after the Big Bang ($z{\gtrsim}11$) during the initial stages of reionization, finding a tantalizingly high abundance of bright, potentially massive galaxy candidates, unanticipated by theoretical models \citep[e.g.,][]{Castellano2022,Naidu2022,Donnan2022,CurtisLake2022,Finkelstein2022,carniani24}. Furthermore, the observed number of massive quenched galaxies exceeds expectations across $z\sim3$--$9$ \citep[][]{carnall22,degraaff24,glazebrook24,narayakkara24}, a number of low-luminosity and dusty quasars \citep{onoue23,harikane23c,matthee23,kocevski23} or active galactic nuclei (AGN)-galaxy complexes \citep{ubler23,larson23,scholtz23,greene23} have been revealed, and some galaxies seem to already have (proto-) disk structures \citep[][]{ferreira22,fudamoto22,guo22}.
However, these early results have come from just a handful of small public legacy fields \citep[e.g.,][]{mcleod24}, leaving open the question of how representative they are of the entire early Universe. More recently, \citet[][]{willott24} extended the analysis to five new fields and found no luminous galaxies at $z>10$, a stark contrast to the early studies. 

Luminous galaxies are rare \citep[$< 10^{-5}$\,Mpc$^{-3}$ for $M_{\rm UV}<-21$ at $z>7$;][]{bouwens15} and inevitably subject to cosmic variance \citep[][]{somerville04,trenti08,robertson10,bhowmick20}. A good example of strong cosmic variance is that \citet{roberts-borsani16} discovered three bright ($m_\mathrm{AB} < 26$) $z>7$ galaxy candidates in just one of the five CANDELS fields, EGS, whereas only one comparably bright candidate was detected from the other four fields. Subsequent follow-up in EGS with HST, ground-based spectroscopy, and JWST has revealed that this region appears overdense at $z\sim7-9$, and contains the majority of Lyman-$\alpha$ emission detected across all legacy fields at $z>7$, likely pointing to a site of significant star formation that reionized early \citep{stark17,tilvi20,larson22,jung22,tang23,Chen2023,Whitler2023}. Such examples clearly demonstrate that a survey of many sightlines, as a supplement to legacy-type surveys with a small number of sightlines, is necessary for unbiased measurements in the early Universe.

\begin{figure*}
\centering
	\includegraphics[width=0.57\textwidth]{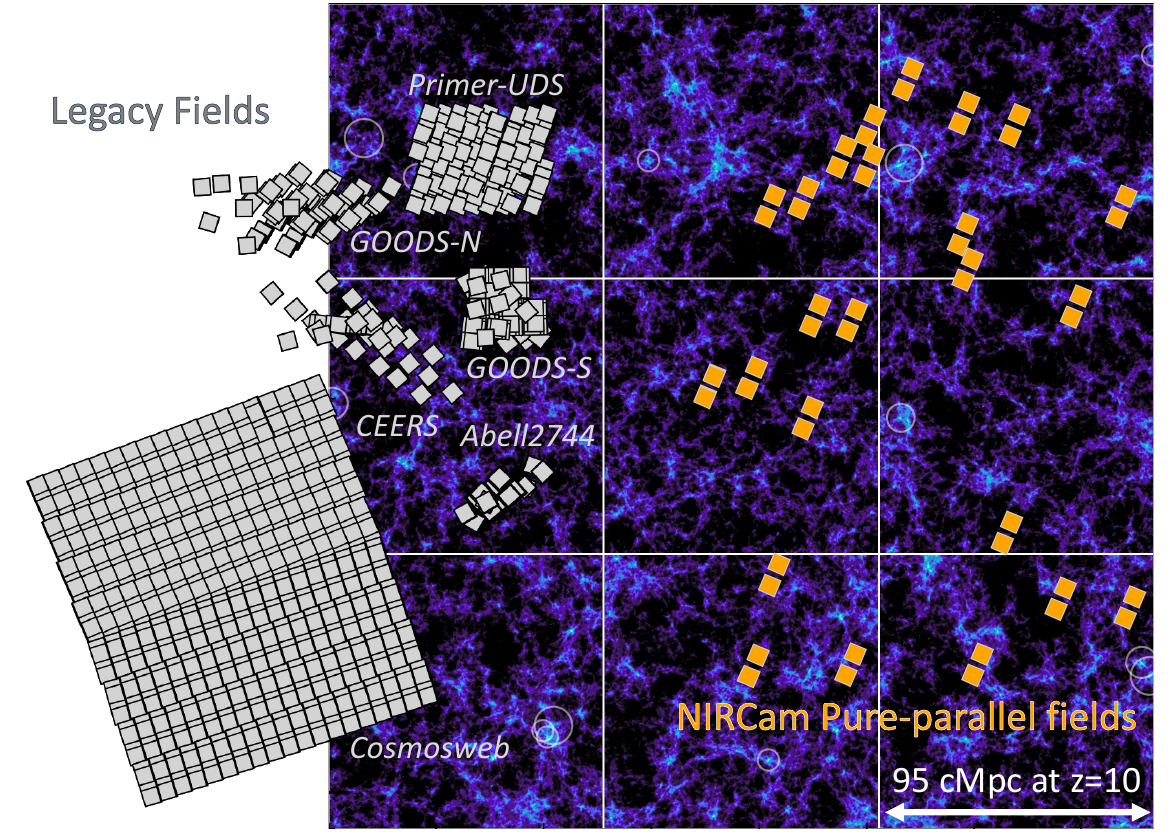}
	\includegraphics[width=0.42\textwidth]{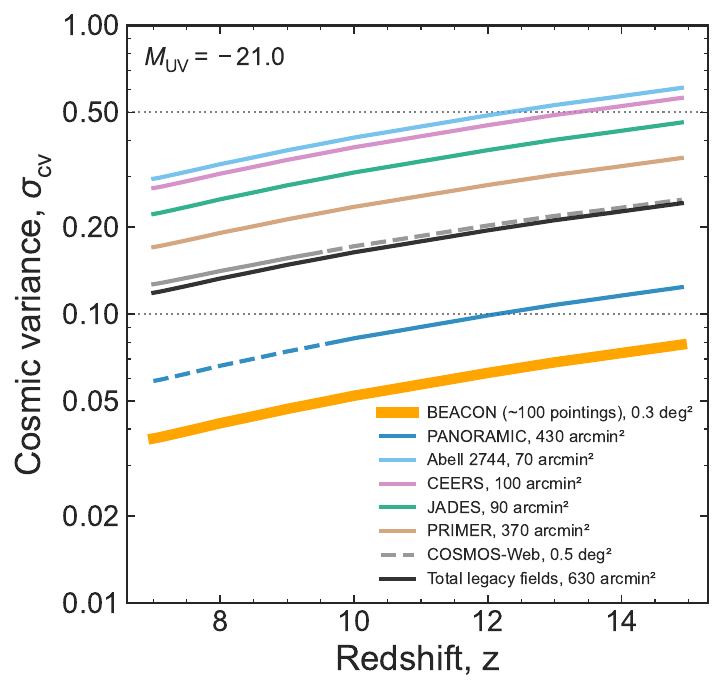}
	\caption{
 (Left): Schematics of NIRCam pure-parallel imaging. Each pair of squares (orange) indicates the size of NIRCam imaging fields, overlaid on dark matter distribution at $z=10$ from the THESAN simulation \citep[][each of the nine panel has a size of $\sim95$\,cMpc]{kannan22}, where halos hosting massive galaxies ($>10^8\,M_\odot$) are encircled. For comparison, Cycle 1 and 2 surveys in legacy fields are shown (gray).  
 (Right): Cosmic variance, $\sigma_\mathrm{cv}$ in the UV luminosity function at $M_\mathrm{UV}=-21$, as a function of redshift for several JWST imaging programs (colored lines), compared to the expected cosmic variance in BEACON (thick orange line), and the total cosmic variance in legacy fields observed with at least 8 NIRCam filters (Abell 2744, JADES, CEERS, PRIMER), all estimated using the cosmic variance calculator by \citet{trapp20} (see Section~\ref{sec:goals}). Dashed lines indicate where contamination is expected to be high in surveys with fewer NIRCam filters.
 }
\label{fig:cv}
\end{figure*}

{\it Pure-parallel} observations offer an ideal opportunity to meet this requirement. While the primary instrument is in operation, a secondary instrument can be used in parallel to observe a field a few arc-minutes away. Since the coordinates of the pure-parallel field cannot be specified, pure-parallel opportunities are often used to identify objects without any prior knowledge, minimizing the effects of cosmic variance \citep[Fig.~\ref{fig:cv}; also][]{atek11,trapp22}.

Starting in 2010, there have been several such HST pure-parallel programs (see \citealt{morishita20} for a review). For example, the BoRG program \citep{trenti11}, which consisted of $>1000$\,\hst\ orbits collected through multiple HST Cycles, successfully identified $z\sim8$--$11$ galaxy candidates at the bright-end magnitude range, $M_{\rm UV}\sim-21$ to $-23$\,mag \citep{trenti12,bradley12,schmidt14,calvi16,bernard16,morishita18b,morishita20}. Identification of such luminous sources from a sufficiently large volume is critical to determine the shape of the luminosity function and its evolution with minimal cosmic variance \citep[][]{bowler14,bowler20,bouwens15,livermore18,bridge19,ren19,Rojas-Ruiz20,leethochawalit23}.

The unexpected detection of many luminous $z>10$ galaxies with JWST demonstrates one reason that pure-parallel efforts are crucial to continue in the era of JWST.
Pure-parallel imaging with NIRCam is ideally suited to robustly identify $z>10$ galaxies and to characterize the shape of UV luminosity functions. Surveying many independent sightlines samples the Universe in an unbiased way and enables studies of galaxy evolution as a function of a wide range of environments \citep[see also][for their Cycle~1 NIRCam Pure-parallel imaging program, PANORAMIC]{williams24}. 

In addition, the pure-parallel mode is essentially at no cost in JWST primary observing time, thus enhancing current and planned programs in legacy fields. Thanks to the dichroic blue and red channels, pure-parallel imaging with NIRCam surveys are twice as efficient as on \hst, and uniquely sample wavelengths of $>2\,\mu$m which is inaccessible to wider-area optical -- near-IR missions like Euclid and Roman.

Identified bright galaxies are ideal targets for follow-up spectroscopy from ground-based facilities \citep{treu13,morishita20,roberts-borsani22borg} and now with JWST. Early NIRSpec follow-up on previously identified \hst\ Borg sources confirmed 10 sources with $<1$\,hr exposure times \citep{roberts-borsani24borg,Rojas-Ruiz24}. However, the low-redshift contamination rate among these HST-selected sources was significant -- in the range $\sim20-50$\,\% depending on redshift, depth of observations and available bands bluewards of the Lyman break \citep{leethochawalit23}. This was expected \citep[e.g.,][]{vulcani17}, and largely attributable to the small number of HST filters up to 1.6\,$\mu$m (i.e., rest-frame $>2000$\,\AA\ at $z>8$) and the lack of sensitive constraints at longer wavelengths that cover rest-frame optical. Multi-band JWST NIRCam imaging greatly improves our ability to obtain robust $z>8$ galaxy samples with a much lower low-$z$ interloper fraction \citep[e.g.,][]{morishita23}.

Here, we introduce Bias-free Extragalactic Analysis for Cosmic Origins with NIRCam (BEACON), a new JWST pure-parallel imaging survey approved for 600 pure-parallel hours in Cycle~2. BEACON is optimized to detect $z>7.3$ Lyman-break galaxies, but also is capable of identifying lower redshift galaxies with up to 8-filter imaging. By surveying $\sim100$ high-latitude blank fields (equivalent to $\sim0.3$\,deg$^2$), it will allow the identification of galaxies over a wide redshift range.

In this overview paper, we present the survey design and initial highlights from the first \nfld\ observed fields. Dedicated studies of UV luminosity functions and galaxy properties will be presented in forthcoming papers (Kreilgaard, in prep.; Zhang, in prep.). In Sec.~\ref{sec:goals}, we outline the science goals of our program. We present the observing configurations and data reduction in Sec.~\ref{sec:data}, followed by photometric analysis of the initial fields in Sec.~\ref{sec:ana}. We release the initial dataset on a dedicated webpage\footnote{\url{https://beacon-jwst.github.io}}.

Where relevant, we adopt the AB magnitude system \citep{oke83,fukugita96}, cosmological parameters of $\Omega_m=0.3$, $\Omega_\Lambda=0.7$, $H_0=70\,\kms\, {\rm Mpc}^{-1}$, and the \citet{chabrier03} initial mass function (IMF).

\section{Survey Goals}\label{sec:goals}

BEACON takes advantage of the unique pure-parallel opportunity provided by the observatory to obtain $1-5\,\mu$m NIRCam imaging over $\sim100$ independent pointings. This enables a wealth of science investigations with minimal cosmic variance. We briefly describe our key science drivers.

\subsection{Unbiased view of galaxies in the first billion years}

Early JWST observations discovered an unprecedented abundance of intrinsically luminous $z{\simgt}10$ galaxy candidates \citep[e.g.,][]{Castellano2022,Naidu2022,Donnan2022,harikane23,Bouwens2022b,casey23}. Those luminous galaxies were not expected by theoretical models \citep[e.g.,][]{dayal14,mason15,mashian16,tacchella18,behroozi18,yung18,Wilkins2022}, implying that some key physics is missing in the model of first galaxy formation. However, these detections have come from just $\sim600$\,sq.~arcmin of public imaging in legacy fields, and are subject to significant cosmic variance. In Fig.~\ref{fig:cv}, we show the fractional cosmic variance as a function of redshift in JWST Cycle 1-2 programs, compared to the predicted cosmic variance in BEACON observations. We estimate the cosmic variance in the UV LF ($\sigma_\mathrm{cv}^2(\phi) = [\langle \phi^2 \rangle - \langle \phi \rangle^2]/\langle \phi \rangle^2$) at $M_\mathrm{UV} = -21$, i.e. comparable to the UV magnitude of the brightest $z>12$ galaxies confirmed by JWST: GHZ2/GLASS-z12 \citep{Castellano2022,Naidu2022} and JADES-GS-z14-0 \citep{carniani24}, using the calculator by \citet{trapp20}. We calculate the cosmic variance in multiple fields by summing the variance in quadrature following \citet{trapp22}. Robustly measuring the bright end of the UV luminosity function (LF) and its evolution at $z>7$ requires \textit{large} samples of $z>7$ sources over a wide luminosity range and sufficient areas, to maximize statistical robustness against cosmic variance. 
Crucially, Fig.~\ref{fig:cv} shows how independent-sightline surveys like BEACON and PANORAMIC \citep{williams24} efficiently reduce cosmic variance compared to contiguous fields. The \citet{trapp20} model predicts that BEACON will reach $\sim5\%$ cosmic variance in the bright-end of the UV LF at $z\sim12$, compared to $\sim20\%$ cosmic variance in both a comparable contiguous area with COSMOS-Web, which has a smaller filter set (4 NIRCam filters), and to the combination of deep JWST surveys in legacy fields (Abell 2744, CEERS, JADES, and PRIMER).

\begin{figure}
\centering
	\includegraphics[width=0.49\textwidth]{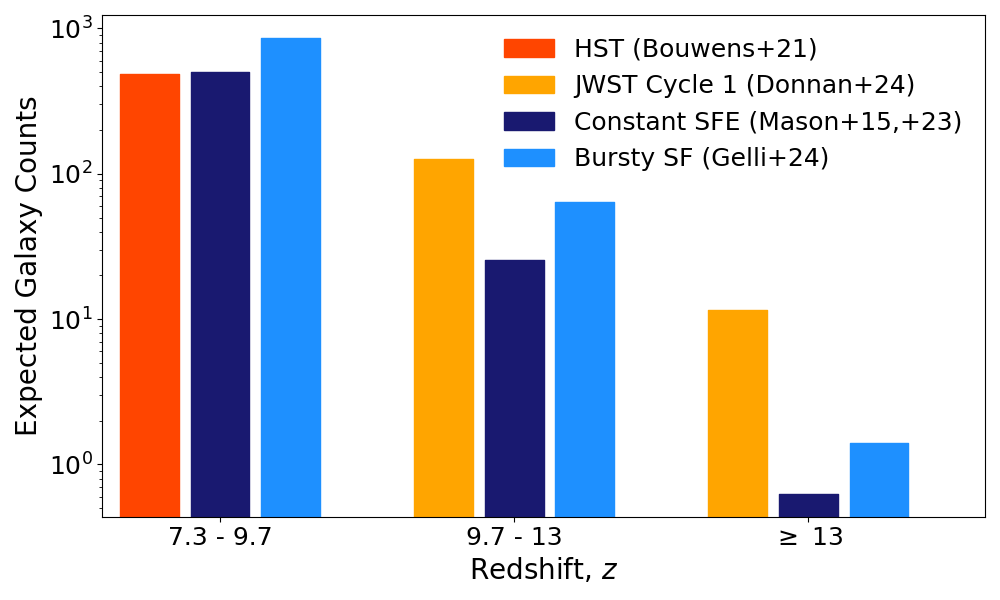} 
	\caption{Expected numbers of sources in BEACON, shown for two theoretical models: constant star-formation efficiency (dark blue, \citealp{mason15, mason22}) and bursty star-formation (light blue, \citealp{gelli24}), and for two data-driven LFs: one determined from HST (red, \citealp{bouwens21}) and one from JWST Cycle 1 (orange, \citealp{donnan24}). All numbers are estimated assuming a flat 80\% completeness down to the 10$\sigma$ limiting magnitude. At $z\geq 13$ models predict hugely different number counts, thus BEACON will distinguish between scenarios.}
\label{fig:ND}
\end{figure}

In Fig.~\ref{fig:ND}, we show the expected number of galaxies that BEACON will find, assuming that the survey will reach 80\% completeness down to $10\,\sigma$ limiting magnitudes. For the allocated 600\,hrs (see also Sec.~\ref{sec:data}), we expect to find 500--1000 galaxies at $z\geq7.3$
and 25--125 galaxies in the redshift range $9.7 \leq z \leq 13$, depending on the model used. The most interesting redshift range is $z\geq13$, where the predicted numbers from different models deviate the most. For example, a model assuming redshift-independent star-formation efficiency even with no dust obscuration \citep{mason15,mason22}, predicts just one galaxy, whereas the LF based on early Cycle 1 fields \citep{donnan24} predicts 12 galaxies and a bursty star formation model \citep{gelli24} does a few. By the completion of our survey, we will have an improved census of galaxies in the first billion years.

\subsection{Cosmic Noon --- Build-up of stellar mass and structures}

Galaxies at $z\sim2$ appear significantly different from those in the local Universe --- with clumpy and disturbed features from star formation \citep[e.g.,][]{elmegreen05,forster-schreiber09,genzel11,guo15}, or compact morphology for post-starburst and passively evolving galaxies \citep[e.g.,][]{daddi05,damjanov09,barro17,setton22}. 
The discovery of already quenched populations at $z>3$ \citep[e.g.,][]{glazebrook17,schreiber18,tanaka19,valentino20}, some of which are clustered \citep{ito23,tanaka24}, suggests a very early and rapid build-up of stellar mass.
Early JWST NIRCam studies identified several massive galaxies at $z\sim4$--7 \citep{carnall22,carnall23,glazebrook23,weibel24}. The prevalence of many evolved galaxies in such early epoch would invoke a critical, but intriguing, need of updates to the current galaxy evolution and quenching models \citep[e.g.,][]{boylan-kolchin22,lovell23}. Further identification requires a larger volume to build a complete picture of massive galaxy build-up. 

BEACON will enable the search for massive galaxies at $2<z<7$, over a total volume of $\sim10^7$\,Mpc$^3$. Based on our magnitude limits and the mass functions from \citet{davidzon17,weaver22}, we expect that $\sim10^5$ galaxies of $M_*>10^9\,M_\odot$ will be identified, including several $\logm>3\times10^{10}\,M_\odot$ galaxies at $z>5$. The star-formation activity of these galaxies will be characterized by, e.g., the rest-frame $UVJ$ diagram \citep[][]{williams09,valentino23}. 
The wavelength coverage from contiguous filters out to $\sim5\,\micron$ will secure galaxy samples of currently missed quiescent and dust-enshrouded optical-dark galaxy populations \citep{wang24,barrufet24}, allowing for a comprehensive study of their properties, including their dependence on galaxy overdensity \citep{morishita24b,champagne24}.
Furthermore, \hst\ pure-parallels demonstrated statistics measured in many uncorrelated fields and provided novel measurements of galaxy/dark matter halo bias \citep{cameron19}. The superb photometric quality will enable searches for candidate over-densities of galaxies \citep[e.g.,][]{newman14}.

\subsection{Nearby Universe --- Dwarfs and galaxies}

Early Release Science (ERS) and Cycle 1 programs clearly showed that a few hours of exposure time with JWST are comparable to the deepest fields observed by HST. BEACON will provide a unique and extensive dataset for legacy science, including enabling follow-up spectroscopy of exciting targets in future JWST cycles. Some legacy science highlights are:

(1) Study of brown dwarfs in the Milky Way, to probe its structure from random sightline
observations \citep{ryan11,holwerda14b,holwerda18}. The wavelength coverage at ${>}1.6\,\mu$m is critical to characterize (sub-)dwarf populations \citep{schneider20,nonino22,burgasser24,hainline24}.

(2) Dust properties of galaxies at $z\sim0$: Stellar continuum and dust emission coexist in the rest frame $1$--$5\,\mu$m. NIRCam filters will enable starburst/AGN diagnostics with F200W$-$F444W color \citep[e.g.,][]{vulcani23} and $3.3\,\mu$m PAH features measured with F356W flux excess \citep[e.g.,][]{inami18} for those with accurate redshift measurements.

\begin{figure}
\centering
	\includegraphics[width=0.45\textwidth]{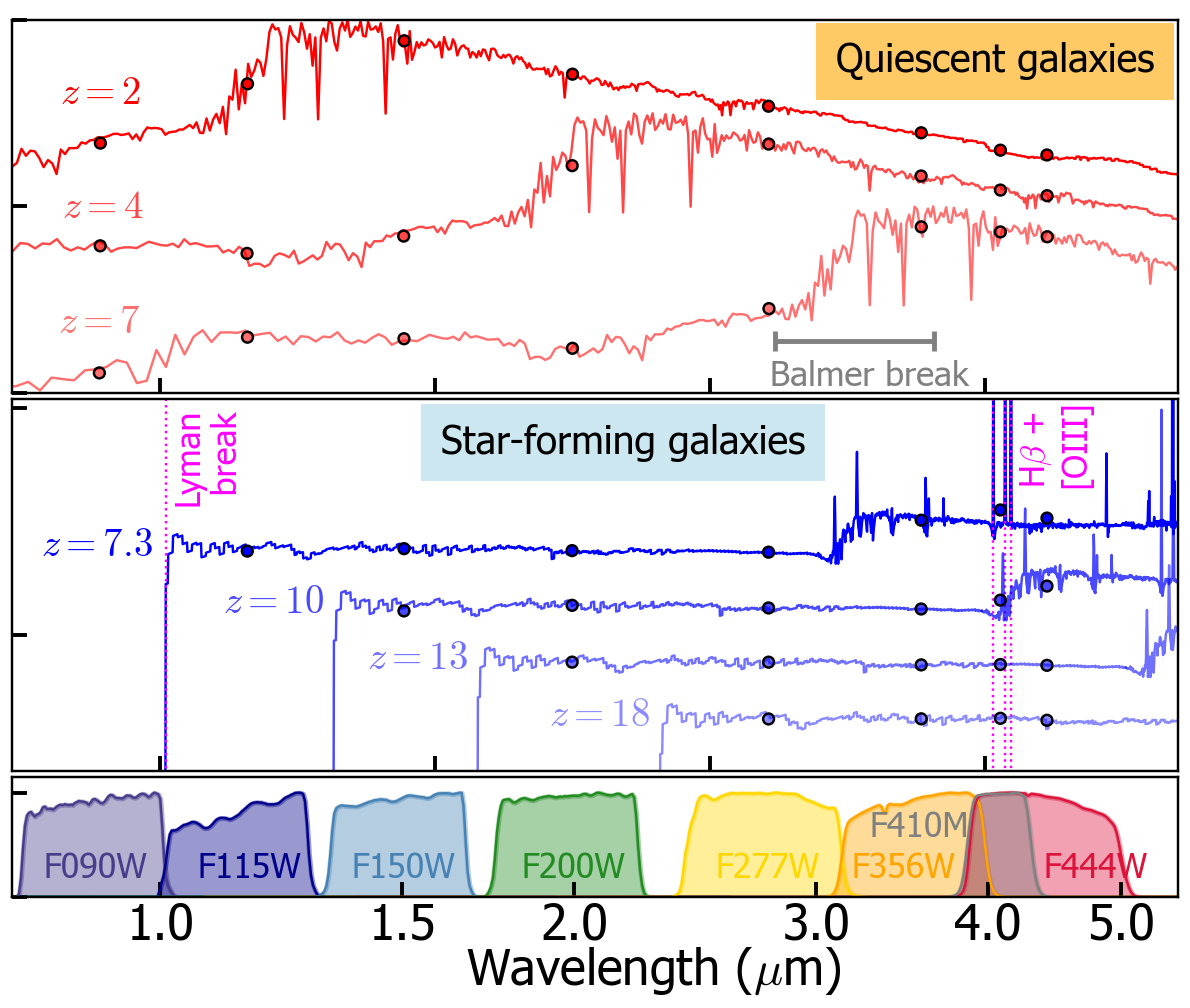}
	\caption{
Template SEDs of galaxies at various redshifts, along with the transmission curves of our default filter set.
    }
\label{fig:filter}
\end{figure}

\section{Survey Design}\label{sec:data}

\subsection{Pure-parallel Observations with JWST}\label{sec:pp}

JWST offers a pure-parallel observing mode\footnote{\url{https://jwst-docs.stsci.edu/jppom/parallel-observations}}, in which one instrument is used while another serves as the primary. This is similar to the pure parallel mode operated by HST \citep{atek11,trenti11,yan11}. With JWST, pure-parallel observations can utilize a variety of instruments (i.e. NIRCam, NIRISS and MIRI as of Cycle~2). The pure-parallel use of NIRCam is analogous to the HST's use of WFC3-IR, as seen in the BoRG survey \citep{trenti11}. However, there are significant differences with NIRCam's pure-parallel imaging: 1) extended red wavelength coverage ($5\,\mu$m vs. $1.7\,\mu$m), 2) a larger effective field-of-view (FoV) per image ($9.7$\,arcmin$^2$ vs. $4.6$\,arcmin$^2$), and 3) dichroic imaging capability, allowing simultaneous short- and long-wavelength observations. The combination of the second and third points makes JWST's NIRCam pure-parallel mode approximately four times more efficient than HST's, even before considering the significant sensitivity differences (i.e. $m_{H}=28.4$\,mag vs 26.6\,mag at $10\,\sigma$, 1\,hr exposure). Additionally, JWST's primary observing modes are typically equipped with dithering, which benefits parallel imaging by reducing potential contamination from detector artifacts and cosmic rays, whereas this has not been often the case for HST's pure-parallel observations.

JWST's pure-parallel mode, however, introduces high complexity, resulting in stringent requirements for scheduling and filter configurations. JWST does not allow mechanical movement during the primary dither, resulting in less flexibility when distributing exposure time across multiple filters in a given pointing. Especially, JWST can offer a very long exposure per visit, whereas HST is limited to a maximum exposure $\sim2800$\,sec due to the interruption from its orbit occultation. This in general helps the primary observations reduce the overhead, but it also means {\it fewer opportunities} for the parallel to reconfigure optical elements. Additionally, the complexity of the spectroscopic observing modes necessitates careful consideration of data rates, which in some cases results in less optimized for artifact rejections.

Among the allocated opportunities, we prioritized those that offer four or more ``group slots'' (corresponding to the number of short-/long-filter pairs) within the same field. As pure-parallel availability is determined by the setup of primary programs on the schedule, not all the allocated opportunities have at least four slots. Thus, we also utilize observations with fewer slots when they either (i)~provide sufficient depth ($\simgt1000$\,sec) or (ii)~include existing NIRCam data covering most of the FoV from previous observations, which can be combined with the new data. We discuss the selection of filters in more detail in Sec.~\ref{sec:filter}.

\subsection{Exposure Configuration}\label{sec:exp}
Each exposure is set to optimize the signal-to-noise ratio (SNR) and meet the data volume limitation. The total data volume is significantly affected by the configuration of each primary observation. As such, a careful consideration for each visit on the Astronomer's Proposal Tool (APT) is required. 
Our default is to use the SHALLOW4 readout mode, which allows a reasonable exposure and SNR without suffering from saturation. When data volume is severely limited, we use a longer readout mode, i.e., DEEP8 or DEEP2, which are more subject to saturation for bright sources. 

\subsection{Filter Configuration}\label{sec:filter}
Taking advantage of NIRCam dichroic, our default filter configuration consists of 8 filters ($0.8 {<} \lambda/\mathrm{\mu m} {<} 5.0$), pairing the four long-wavelength filters (F277W, F356W, F410M, F444W) with four short-wavelength ones (F090W, F115W, F150W, F200W; Fig.~\ref{fig:filter}).  This means that we require a minimum of four group slots in the primary program, where each group slot consists of an entire dither sequence. However, due to the limited availability of parallel slots, our program utilizes opportunities of less than 4 group slots (see below).

Those 8 NIRCam filters securely capture the Lyman break of galaxies at $z{>}7$ and the Balmer break for galaxies at $2<z<7$. The almost uninterrupted, non-overlapping filter choice at $0.8$\,-\,$5\,\mu$m allows us a complete sampling of galaxies at these redshift ranges. In addition to the Lyman break, the Balmer break sampled by F277W/F356W/F410M/F444W is critical to characterize stellar populations of galaxies at $z \simlt10.5$ \citep[e.g.,][]{witten24,katz24}, as well as to capture extreme line emission from \hb+\oiii\ at $7<z<9$ by measuring flux excess in F410M/F444W \citep[e.g.,][]{Endsley2022,llerena24}. 
When five or more group slots are available, we include other medium-band filters (F140M, F182M, F210M, F430M, F460M) which increase the redshift reliability and expand the redshift range of strong-line emitter searches.

For parallel opportunities where only three group slots are available, we configure those with F090W, F115W, F150W, F277W, F356W, and F444W. This strategy keeps a redshift range similar to that for the full filter case. We utilize opportunities with fewer filters only when there are available data in the overlapping area; for those, we select unused filters to increase wavelength sampling.

We note that several visits so far were skipped in our executed observations, due to scheduling issues (unexpected changes in the primary observations) or data transfer limits during the visits. Consequently, those affected fields may have shorter total exposures than planned, or only partial filter coverage, or in some cases no observations.

\section{Data analysis overview}\label{sec:ana}

\subsection{Imaging Data Reduction}\label{sec:data-jwst}
\begin{figure*}
\centering
	\includegraphics[width=0.95\textwidth]{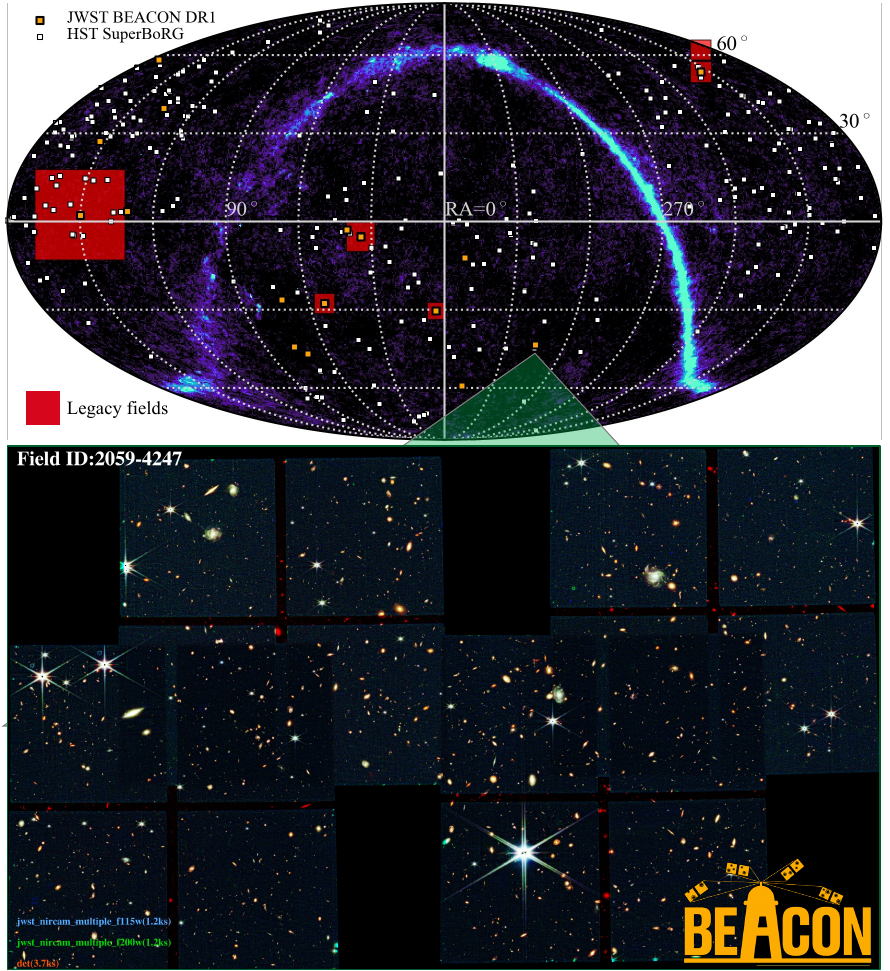}
	\caption{
 ($Top$): Sky distributions of the \nfld\ BEACON DR1 fields (orange squares), overlaid on a temperature map from the WMAP 7-yr data \citep{jarosik11}.
 Also shown are legacy fields of JWST (COSMOS-Web, PRIMER-UDS, CEERS-EGS, JADES-GOODS-South, JADES-GOODS-North, and Abell2744; red squares) and the HST's pure-parallel imaging fields from HST SuperBoRG (white squares; $N=316$). The symbol sizes are not the actual sky size, but are roughly scaled to the corresponding survey area. 
 ($Bottom$): RGB-color composite image of one of the DR1 fields. 
    }
\label{fig:sky}
\end{figure*}

In this study, we reduce and analyze data taken in the first \nfld\ fields. The sky distributions of the fields are shown in Fig.~\ref{fig:sky}. We follow an approach similar to that presented by \citet{morishita23} for the reduction of images. Briefly, our pipeline retrieves the raw-level images from the Mikulski Archive for Space Telescopes (MAST) archive when they become available after each observation, and then reduces them with the official JWST pipeline. Our pipeline adds several custom steps including $1/f$-noise subtraction using {\tt bbpn}\footnote{\url{https://github.com/mtakahiro/bbpn}}, snowball masking using {\tt Grizli} \citep{brammer22}, and additional cosmic-ray masking using {\tt lacosmic} \citep{vandokkum01,bradley23}. The final drizzled images, with the pixel scale set to $0.\!''0315$\,/\,pixel, are aligned to the IR-detection image (F277W+F356W+F444W as the default choice).

To homogenize the PSF of the NIRCam images with that of F444W, we generate convolution kernels using PSFs generated with {\tt webbpsf}\footnote{\url{https://github.com/spacetelescope/webbpsf}}. We provide those PSFs to {\tt pypher} \citep{boucaud16} and generate convolution kernels for each filter. The quality of convolved PSFs is within $\ll 1\,\%$ agreement in the encircled flux at radius $r=0.\!''16$ i.e. the size used for our photometric flux measurements.

\begin{figure*}
\centering
	\includegraphics[width=0.95\textwidth]{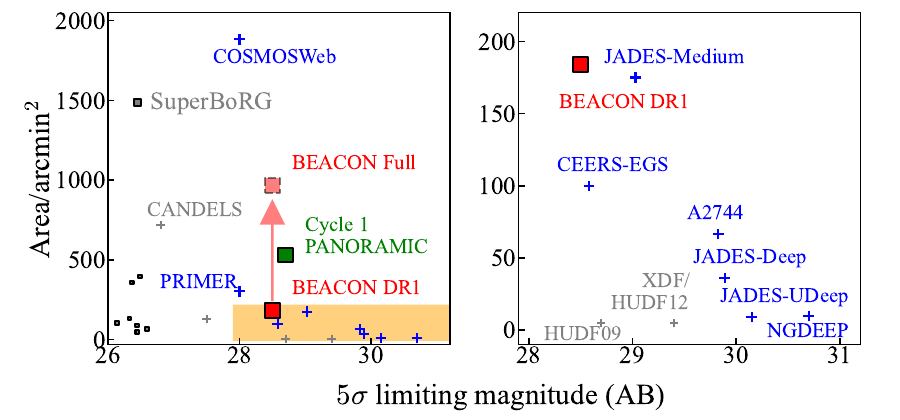}
	\caption{
    Limiting magnitudes and survey areas of various JWST programs (symbols in color). Square and cross symbols represent pure-parallel and for legacy surveys, respectively. Several HST programs are also shown in gray (see \citealt{morishita21} for the full description). The region hatched in orange is shown in zoom in the right panel.
    }
\label{fig:magarea}
\end{figure*}

\begin{figure*}
\centering
	\includegraphics[width=0.95\textwidth]{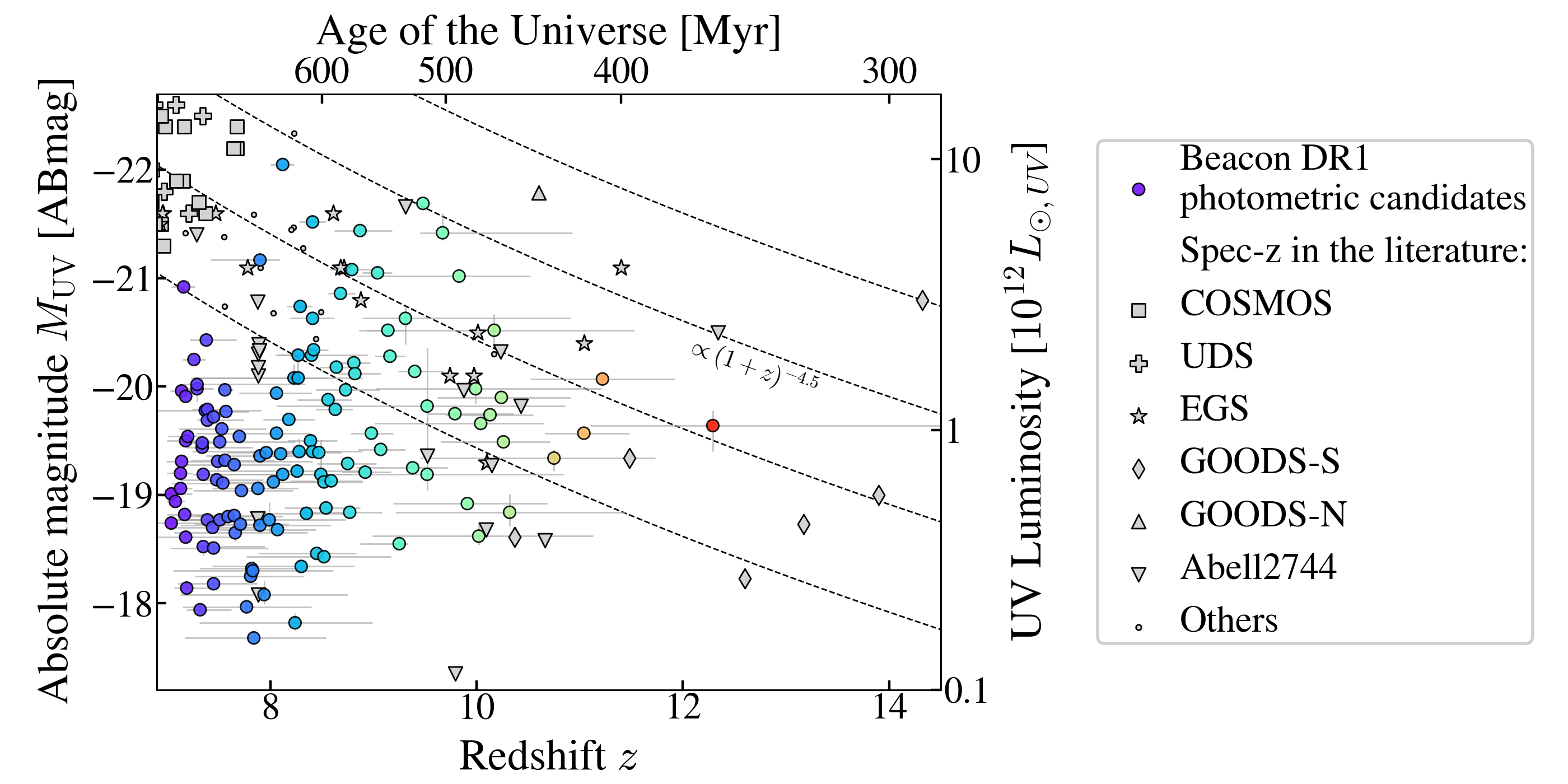}
	\caption{
 Redshift-$M_{\rm UV}$ distributions of final photometric candidates selected from the \nfld\ DR1 fields (circles, colored by redshift). Semi-empirical luminosity evolution curves,  $\propto(1+z)^{-4.5}$, of halos of a given comoving abundance are shown (dashed lines). Spectroscopically confirmed sources in the literature are shown (gray symbols; \citealt{curtis-lake23,harikane23c,arrabalharo23b,morishita23b,castellano24,roberts-borsani24,napolitano24,carniani24}).
    }
\label{fig:zMUV}
\end{figure*}

\begin{deluxetable*}{ccccccccccccccccccc}
\tabletypesize{\footnotesize}
\tablecolumns{18}
\tablewidth{0pt}
\tablecaption{
}
\tablehead{
\colhead{FLDID} & \colhead{R.A.} & \colhead{Decl.} & \multicolumn{2}{c}{F090W} & \multicolumn{2}{c}{F115W} & \multicolumn{2}{c}{F150W} & \multicolumn{2}{c}{F200W} & \multicolumn{2}{c}{F277W} & \multicolumn{2}{c}{F356W} & \multicolumn{2}{c}{F410M} & \multicolumn{2}{c}{F444W}\\
\cline{4-5}
\cline{6-7}
\cline{8-9}
\cline{10-11}
\cline{12-13}
\cline{14-15}
\cline{16-17}
\cline{18-19}
\colhead{} & \colhead{deg.} & \colhead{deg.} & \colhead{${t_{\rm exp}}$} & \colhead{${m_{5\sigma}}$} & \colhead{${t_{\rm exp}}$} & \colhead{${m_{5\sigma}}$} & \colhead{${t_{\rm exp}}$} & \colhead{${m_{5\sigma}}$} & \colhead{${t_{\rm exp}}$} & \colhead{${m_{5\sigma}}$} & \colhead{${t_{\rm exp}}$} & \colhead{${m_{5\sigma}}$} & \colhead{${t_{\rm exp}}$} & \colhead{${m_{5\sigma}}$} & \colhead{${t_{\rm exp}}$} & \colhead{${m_{5\sigma}}$} & \colhead{${t_{\rm exp}}$} & \colhead{${m_{5\sigma}}$}
}
\startdata
0014-3025 &3.59520 &-30.41974 &10.0 &28.9 &10.0 &28.8 &-- &-- &10.0 &29.1 &10.0 &29.1 &10.0 &29.3 &-- &-- &10.0 &29.0 \\
0217-0509 &34.21885 &-5.13408 &18.9 &29.3 &13.1 &29.0 &13.1 &29.2 &-- &-- &13.1 &29.1 &13.1 &29.2 &-- &-- &18.9 &28.8 \\
0217-0508 &34.29253 &-5.12272 &18.9 &29.4 &13.1 &29.0 &13.1 &29.2 &-- &-- &13.1 &29.2 &13.1 &29.2 &-- &-- &18.9 &28.9 \\
0217-0504 &34.30829 &-5.07792 &18.9 &29.2 &13.1 &29.0 &13.1 &29.2 &-- &-- &13.1 &29.5 &13.1 &29.6 &-- &-- &18.9 &29.3 \\
0240-0253 &40.09595 &-2.87019 &1.0 &26.9 &1.0 &26.9 &1.0 &27.2 &-- &-- &1.0 &28.0 &1.0 &28.1 &-- &-- &1.0 &27.5 \\
0332-2749 &53.03300 &-27.81392 &16.1 &29.4 &16.1 &29.4 &1.2 &28.0 &16.1 &29.7 &16.1 &29.4 &1.2 &28.5 &16.1 &29.1 &16.1 &29.3 \\
0332-2745 &53.03766 &-27.75151 &7.5 &28.9 &7.5 &29.0 &7.5 &29.1 &12.5 &29.3 &7.5 &29.6 &7.5 &29.5 &-- &-- &20.0 &29.2 \\
0442-4613 &70.42773 &-46.21713 &0.7 &26.6 &0.5 &26.1 &0.5 &26.4 &-- &-- &0.5 &27.5 &0.5 &27.5 &-- &-- &0.7 &27.7 \\
0447-2637 &71.67559 &-26.60116 &1.2 &27.6 &1.2 &27.4 &1.2 &27.9 &1.2 &28.1 &1.2 &28.6 &1.2 &28.5 &1.2 &27.9 &1.2 &28.2 \\
0502-4338 &75.45100 &-43.63230 &-- &-- &1.0 &27.5 &1.0 &27.8 &1.0 &28.0 &1.0 &28.5 &1.0 &28.5 &-- &-- &1.0 &28.2 \\
0843+0324 &130.65348 &3.41877 &-- &-- &0.8 &27.6 &0.8 &27.8 &0.8 &27.9 &0.8 &28.3 &0.8 &28.3 &-- &-- &0.8 &27.8 \\
0860+3857 &134.98519 &38.96236 &2.5 &27.9 &1.6 &27.5 &1.6 &27.7 &-- &-- &1.6 &28.5 &1.6 &28.5 &-- &-- &2.5 &28.2 \\
0959+0200 &149.84149 &2.01664 &3.3 &28.3 &3.3 &28.3 &3.3 &28.5 &3.3 &28.6 &3.3 &28.9 &3.3 &28.9 &3.3 &28.2 &3.3 &28.5 \\
1010+2701 &152.44541 &27.03744 &4.2 &28.1 &4.2 &28.1 &4.2 &28.4 &-- &-- &4.2 &28.7 &4.2 &28.7 &-- &-- &4.2 &28.3 \\
1138+5748 &174.41601 &57.80085 &1.4 &27.4 &1.4 &27.3 &1.4 &27.6 &-- &-- &1.4 &28.5 &1.4 &28.4 &-- &-- &1.4 &28.1 \\
1420+5252 &215.07468 &52.86999 &3.4 &28.0 &3.4 &28.1 &3.4 &28.3 &3.4 &28.5 &3.4 &28.9 &3.4 &28.9 &3.4 &28.2 &3.4 &28.6 \\
2059-4247 &314.63703 &-42.78990 &-- &-- &1.2 &27.0 &1.2 &27.3 &1.2 &27.4 &1.2 &28.0 &1.2 &28.1 &-- &-- &1.2 &27.6 \\
2316-5910 &349.11637 &-59.15965 &-- &-- &1.8 &27.0 &1.8 &27.3 &1.8 &27.5 &1.8 &28.1 &1.8 &28.2 &-- &-- &1.8 &27.8 \\
2325-1216 &351.35492 &-12.26333 &-- &-- &1.0 &26.5 &1.0 &26.8 &1.0 &26.8 &1.0 &27.7 &1.0 &27.8 &-- &-- &1.0 &27.6 \\
\enddata
\tablecomments{
5-$\sigma$ limiting magnitudes for point sources are shown. Exposure times are shown in the units of 1000\,seconds.
}\label{tab:fld}
\end{deluxetable*}

\subsection{Photometry}\label{sec:data-jwst}
We identify sources in the detection image using \sext\ \citep{bertin96}. We measure source fluxes on point spread function (PSF)-matched mosaic images with a fixed aperture of radius $0.\!''16$. We set the configuration parameters of \sext\ as follows: DETECT\_MINAREA 0.0081\,arcsec$^2$, DETECT\_THRESH 1.0, DEBLEND\_NTHRESH 64, DEBLEND\_MINCONT 0.0001, BACK\_SIZE 128, and BACK\_FILTSIZE 5.

The aperture-based fluxes measured for each source are then scaled by a single factor, defined as ${f_{\rm auto, F444W}/f_{\rm aper, F444W}}$, to the total flux, where ${f_{\rm auto}}$ is the flux measured within elliptical Kron apertures with the 2.5 scaling factor. With this approach, colors remain as those measured in the aperture, whereas the fluxes used in the following analyses (for the derivation of stellar mass and star formation rate) are scaled up to the total quantities. 

Once the fluxes are measured and scaled, Galactic dust reddening is corrected using the attenuation value retrieved for the coordinates of each field from NED \citep{schlegel98,schlafly11}. We adopt the canonical Milky Way dust law for the reddening curve \citep{cardelli89}.

The average limiting magnitudes of the images are measured in regions where no source is detected among the effective area (i.e. sky area), using apertures of $0.\!''16$ radius, and reported in Table~\ref{tab:fld}. The average limiting magnitude of the DR1 fields and the total area are shown in Fig.~\ref{fig:magarea}, together with those of the aforementioned JWST surveys and several HST surveys \citep{koekemoer11,grogin11,ellis13,illingworth13,bouwens19}.

\subsection{Photometric Redshift}\label{sec:phot-z}
For each field, we produce a photometric catalog by running {\tt eazy} \citep{brammer08}. Our default setup is similar to what was presented in \citet{morishita23b}, using a template library provided by \citet{hainline23}. This template library supplements {\tt eazy}'s default library (v1.3) by adding young, line-emitting galaxy templates generated with {\tt fsps} \citep{conroy09fsps}. The increased variety of fitting templates thus enables a more comprehensive redshift estimate for target young galaxies. The redshift range of fitting is set to $0<z<30$, with a step size of $0.01 (1+z)$. We adopt a default magnitude prior provided by {\tt eazy}, using one of the three NIRCam red channel broad filters (F277W, F356W, F444W) based on their availability in each field.

We utilize the output likelihood function derived by {\tt eazy} to estimate a statistical redshift range for each source. We adopt the median redshift calculated using the full $p(z)$ distribution as the best-fit redshift, and estimate $1\,\sigma$ uncertainties by taking the differences from the 16th and 84th percentile redshifts. The redshift measurements and the probability distributions are used in our high-$z$ galaxy selection (Sec.~\ref{sec:highz}). Absolute UV magnitudes ($M_{\rm UV}$) are calculated using the best-fit template at the corresponding redshifts.

\subsection{Photometric Selection of High-Redshift Galaxies}
We identify candidate galaxies at $z>7$ via the Lyman-break dropout method \citep{steidel98}. Following a similar approach presented in \citet{morishita23,morishita24}, we first apply the following color selection:

{\bf F090W-dropouts ($7.3\simlt z \simlt 9.7$)}
$$S/N_{\rm 150} > 4 $$
$$S/N_{\rm 090} < 2$$
$$z_{\rm set} = 6$$

{\bf F115W-dropouts ($9.7\simlt z \simlt13$)}
$$S/N_{\rm 200} > 4 $$
$$S/N_{\rm 115, 090} < 2$$
$$z_{\rm set} = 8$$

{\bf F150W-dropouts ($13\simlt z \simlt18$)}
$$S/N_{\rm 277} > 4 $$
$$S/N_{\rm 150, 115, 090} < 2$$
$$z_{\rm set} = 10$$
where available filters that are bluer than the rest-frame wavelength of $<1216\,{\rm \AA}$ are used for $2\,\sigma$ non-detection. Furthermore, to secure non-detection, we repeat the non-detection step with a smaller aperture, $r=0.\!''08$ ($\sim2.5$\,pixel). Note that a photometric selection is not attempted for redshift windows where no dropout filter is available. $z_{\rm set}$ is a redshift limit used to calculate total low-$z$ probabilities (see below).

After securing non-detection in blue filters, we apply a photometric redshift selection. Following \citet{morishita18b}, in each dropout selection we exclude sources that do not satisfy $p(z>z_{\rm set})>0.8$, i.e. total redshift probability at $z>z_{\rm set}$ is less than $80\,\%$, where $z_{\rm set}$ is defined above separately for each selection redshift window. The combination of the detection/non-detection requirement and the phot-$z$ analysis provides us with sources whose phot-$z$ solution is consistent with the dedicated redshift range of each dropout selection. 
\section{Initial Results}\label{sec:res}

\subsection{Overview of the Observed Fields} 
\label{sec:ov}
In this overview paper,  we focus on the first \nfld\ fields taken (hereafter Data Release 1, or DR1, fields). Results from the full fields will be presented upon completion of the program in forthcoming papers. In Tab.~\ref{tab:fld}, we summarize the DR1 fields. Limiting magnitudes, calculated for a fixed aperture size, are listed.

Some fields overlap with previous extragalactic JWST fields, including Abell~2744 \citep{Treu22,labbe23b}, COSMOS \citep{casey22}, AEGIS \citep{bagley23ceers}, UDS \citep{donnan24}, and GOODS-South/North \citep{hainline23}. As mentioned previously, in those overlapping fields we supplement the filters that were not previously used (especially F090W; and when possible use medium filters such as F140M/F182M/F210M/F430M/F480M), to enable searches of emission line galaxies at $7<z<9$ and improve photometric-redshift accuracy.

\subsection{Identification of $z>7$ galaxies}\label{sec:highz}
From the DR1 fields, we have identified \nsrc\ galaxy candidates at $z>7$, consisting of \ny\ F090W-dropouts, \nj\ F115W-dropouts, and zero F150W-dropouts. In Fig.~\ref{fig:zMUV}, we show the redshift-$M_{\rm UV}$ distribution of the selected sources. The faint end of the UV magnitude probed here is $\sim-18$,mag, consistent with the prediction by the exposure time calculator. We observe several luminous ($<-21$\,mag) galaxies at $8<z<10$. As we see below (Sec.~\ref{sec:nd}), the observed number of those luminous galaxies is overall consistent with pre-JWST studies \citep[e.g.,][]{bouwens15,stefanon19}, post-JWST \citep{leung23,willott24}, and also theoretical predictions \citep[e.g.,][]{Mason2022}. 

On the other hand, at $z>10$ we do not find many luminous galaxies that are comparable to GN-z11 \citep{oesch18,bunker23,tacchella23}, GHZ2 \citep{Castellano2022,Naidu2022}, and JADES-GS-z14-0 \citep{carniani24}. Also shown in Fig.~\ref{fig:zMUV} are spec-$z$ confirmed sources in the literature. Interestingly, while the total area of the Beacon DR1 is comparable to those in the literature, we seem to be lacking luminous ($<-20$\,mag) galaxies at $z>10$, casting a doubt on our selection being too conservative. However, we confirmed that our color selection would successfully reproduce 12 of the 13 sources at $z_{\rm spec}>10$. The only unsuccessful case (GS-z10-0; \citealt{curtis-lake23}) is a faint galaxy ($m_{200}\sim29$\,mag) and thus our phot-z estimate has large error bars ($z_{\rm phot}=2.5_{-2.2}^{+8.8}$ at $1\,\sigma$). Given the high success rate of our color selection, we suspect that the difference in $z>10$ galaxy number densities may originate in cosmic variance. In fact, 6 and 3 of the $z_{\rm spec}>10$ sources are located in the GOODS-South field and {\it one} CEERS pointing, respectively. A further investigation of cosmic variance will be presented in our future work (Kreilgaard, in prep).

\subsubsection{Overview of Individual High-$z$ Galaxy Candidates}\label{sec:cand}
\begin{figure*}
\centering
	\includegraphics[width=0.48\textwidth]{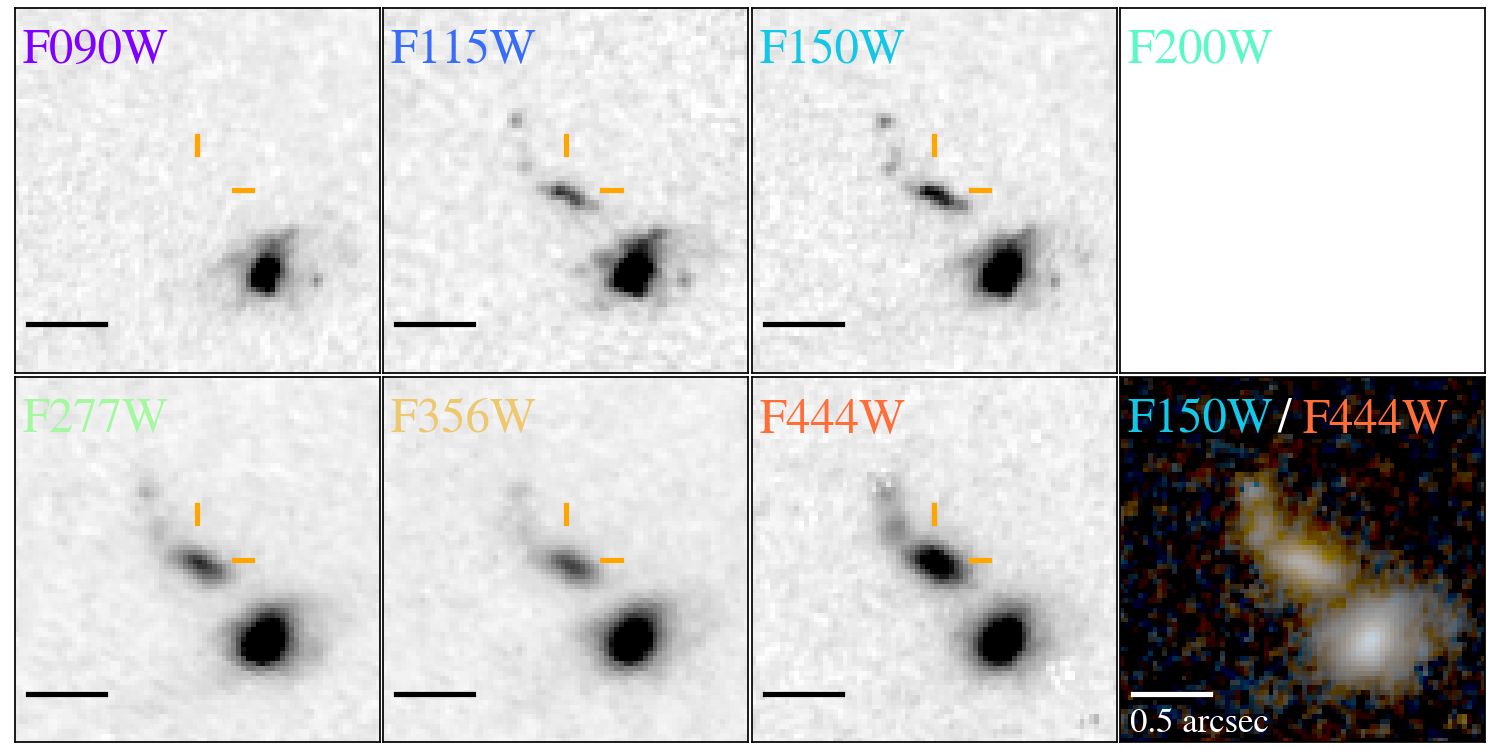}
	\includegraphics[width=0.48\textwidth]{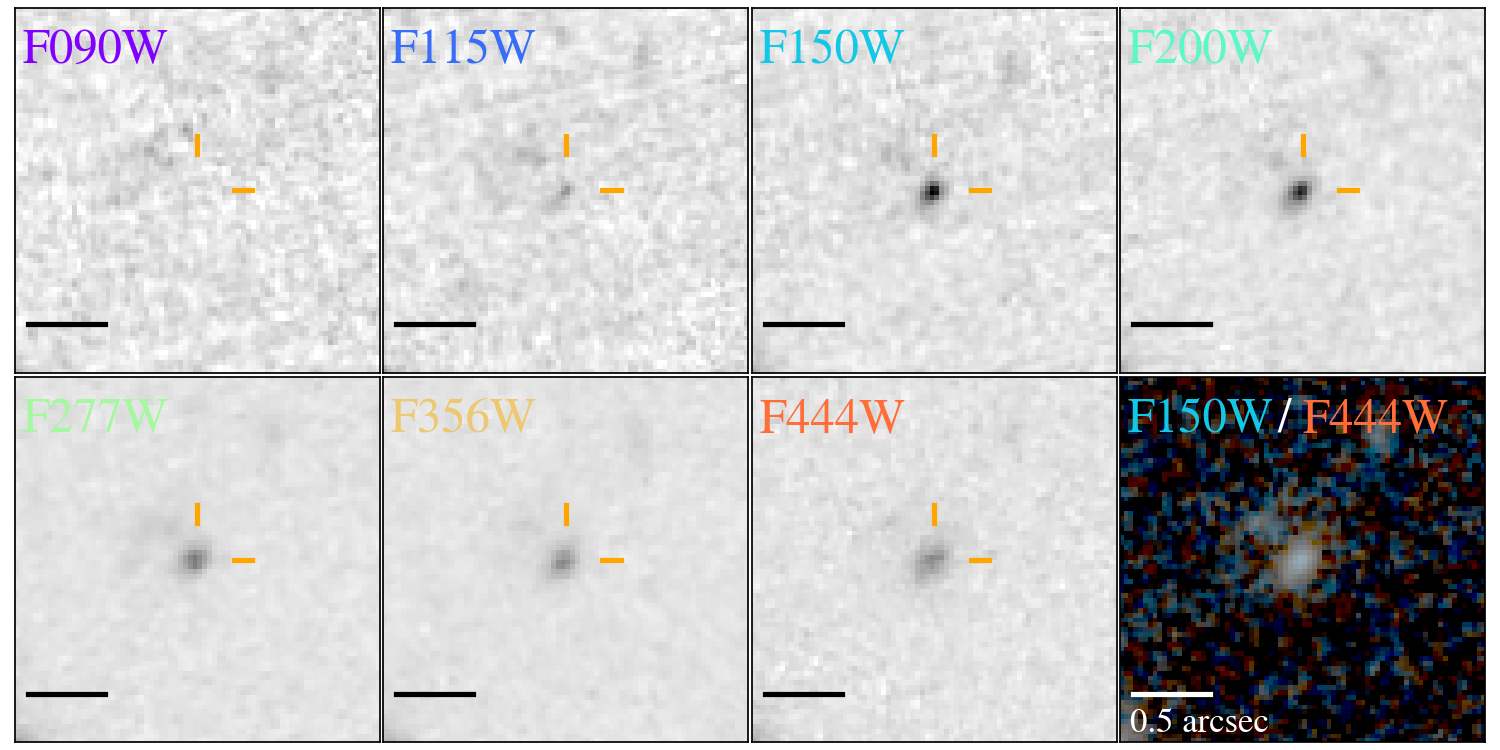}
	\includegraphics[width=0.48\textwidth]{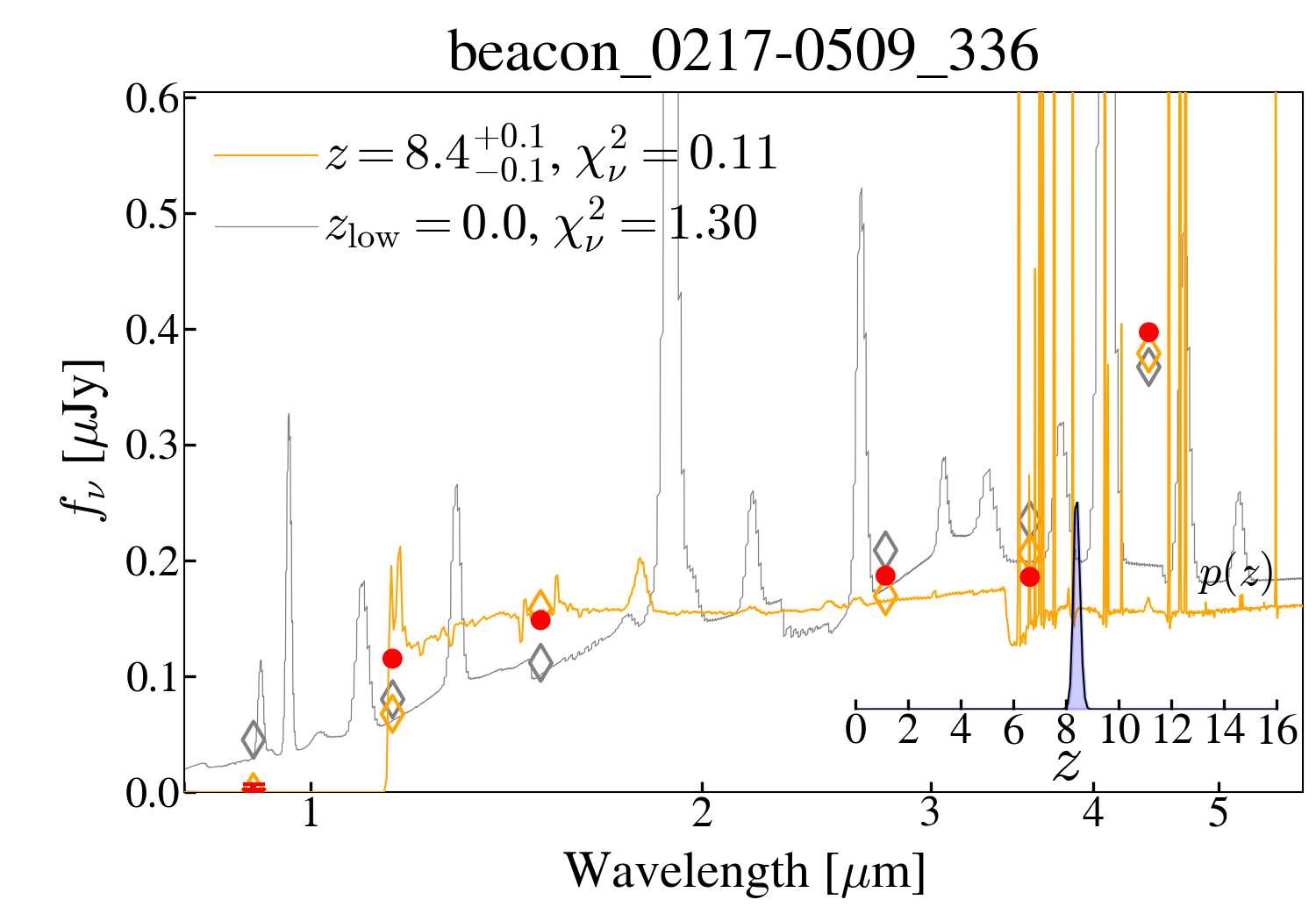}
	\includegraphics[width=0.48\textwidth]{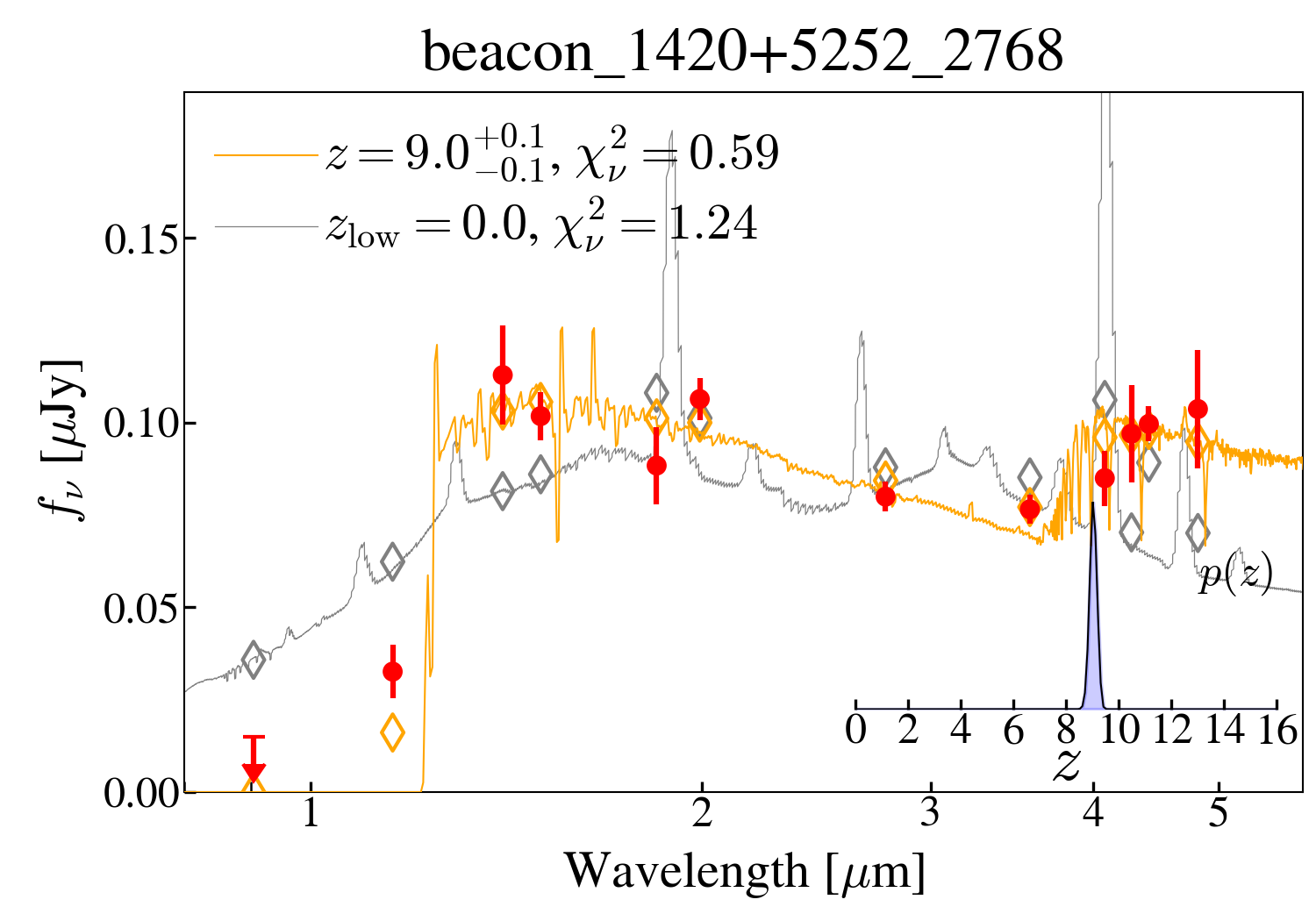}
	\includegraphics[width=0.48\textwidth]{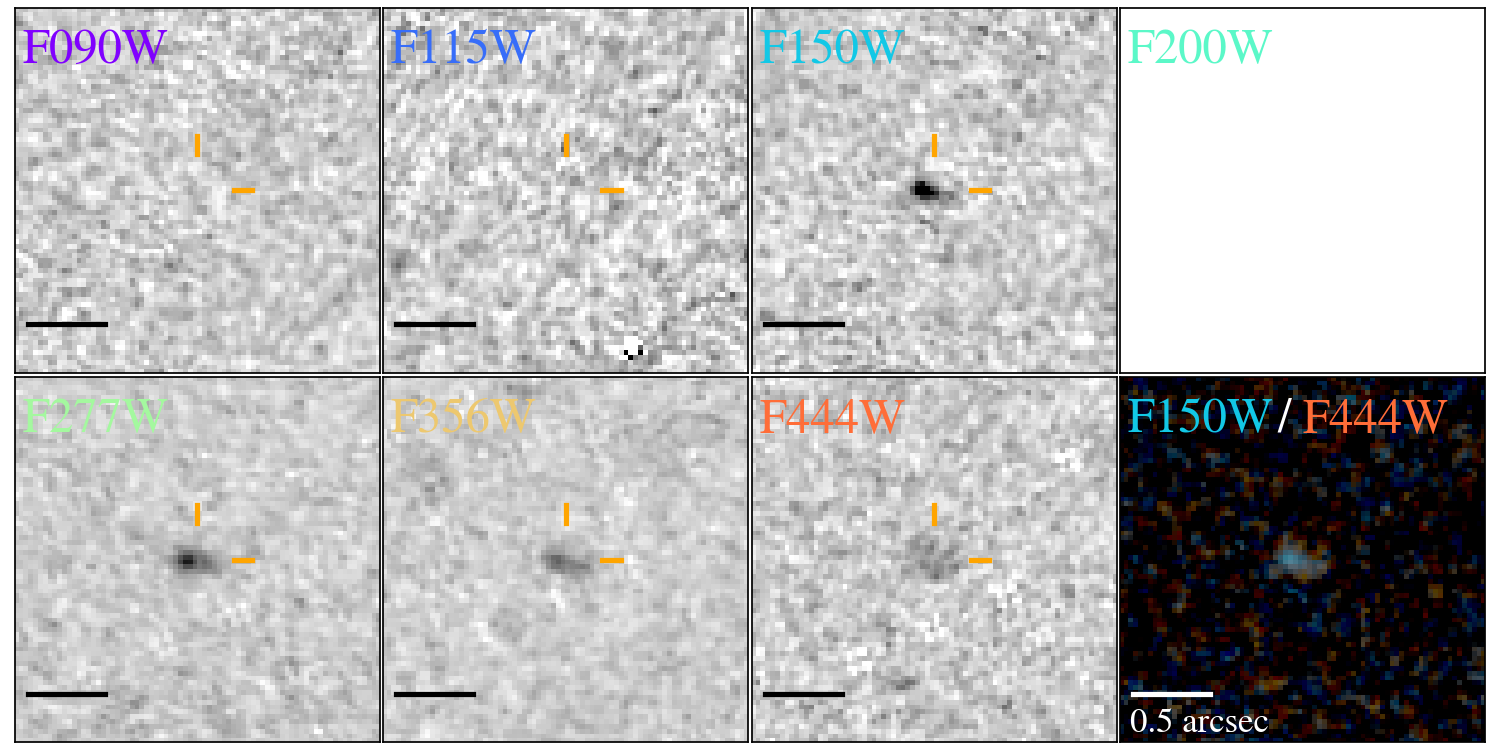}
	\includegraphics[width=0.48\textwidth]{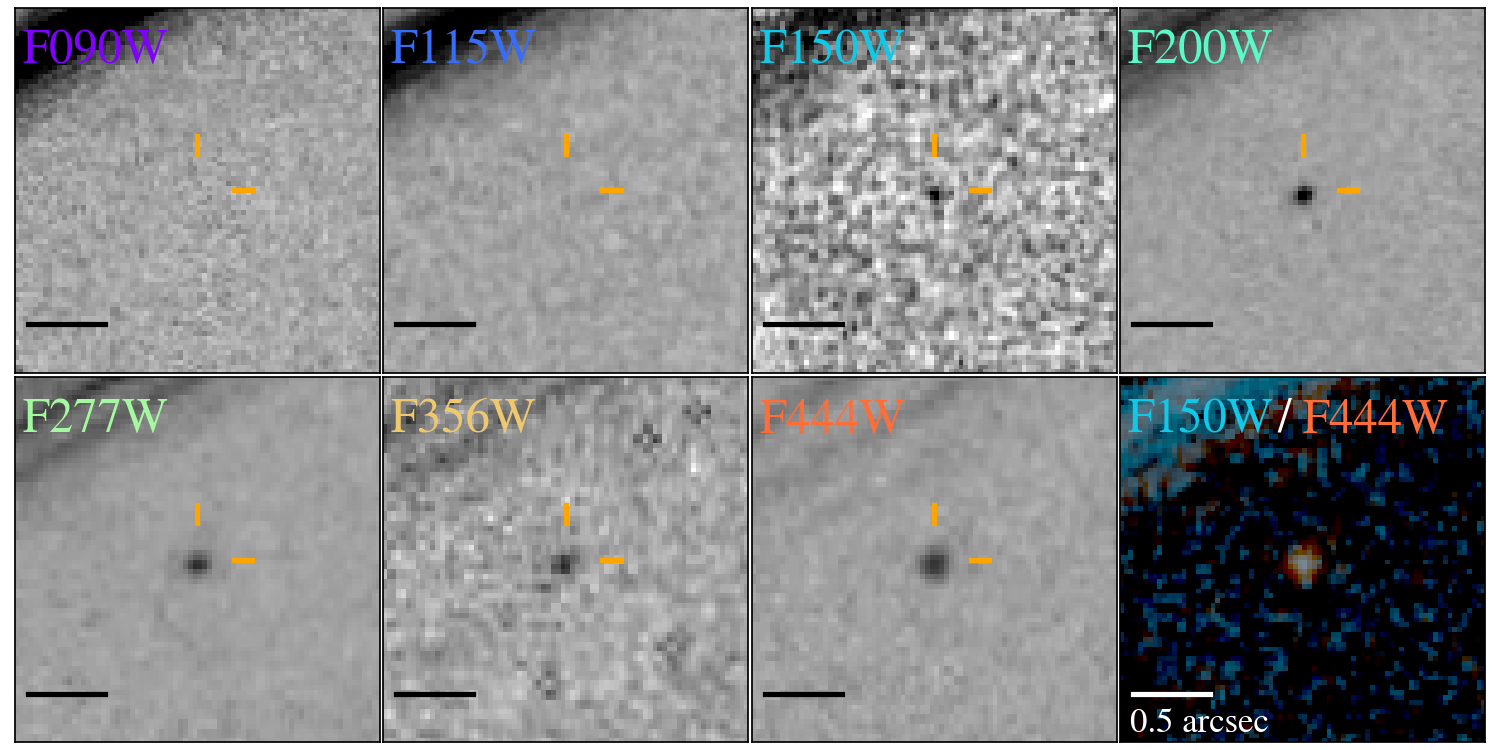}
	\includegraphics[width=0.48\textwidth]{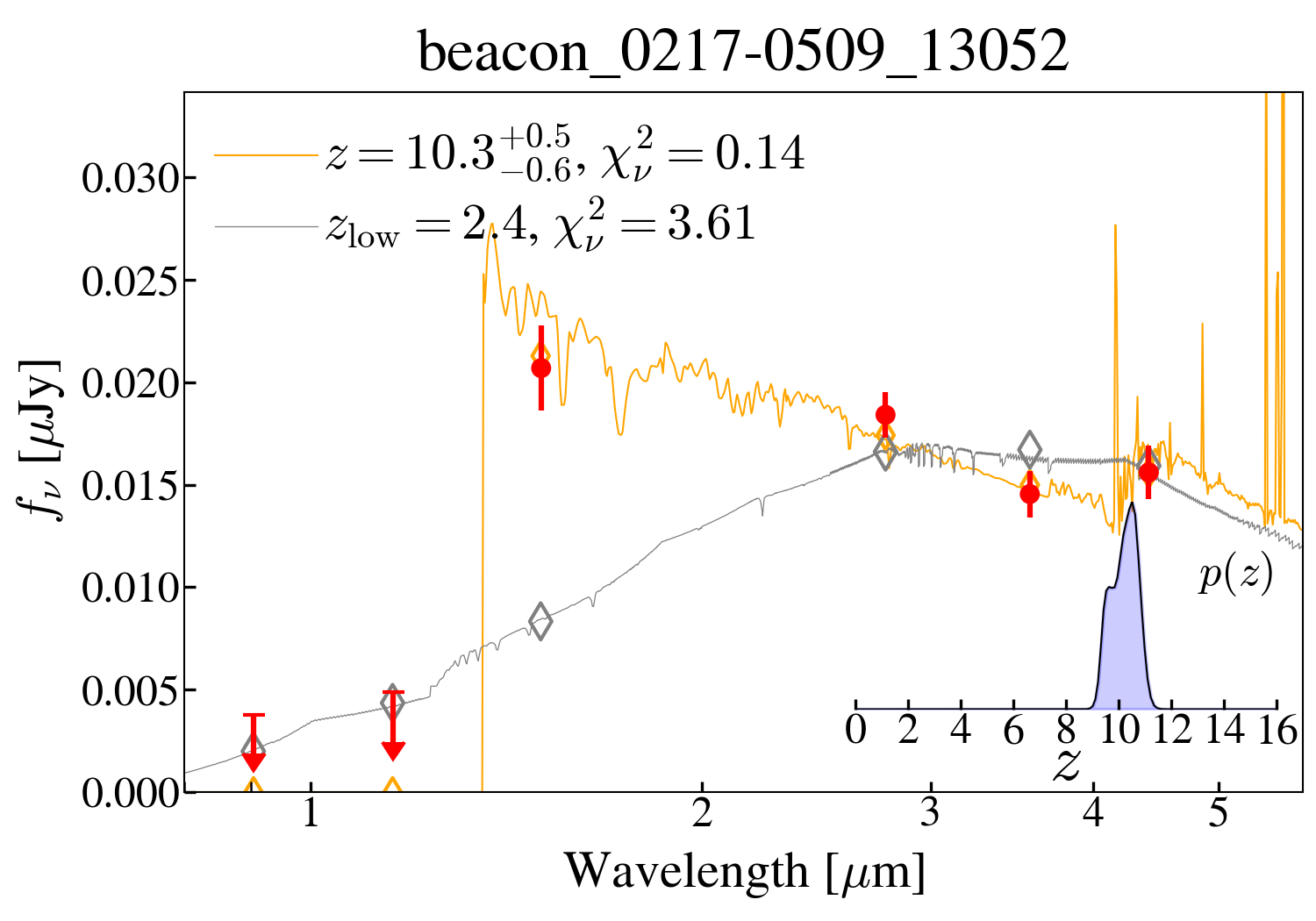}
	\includegraphics[width=0.48\textwidth]{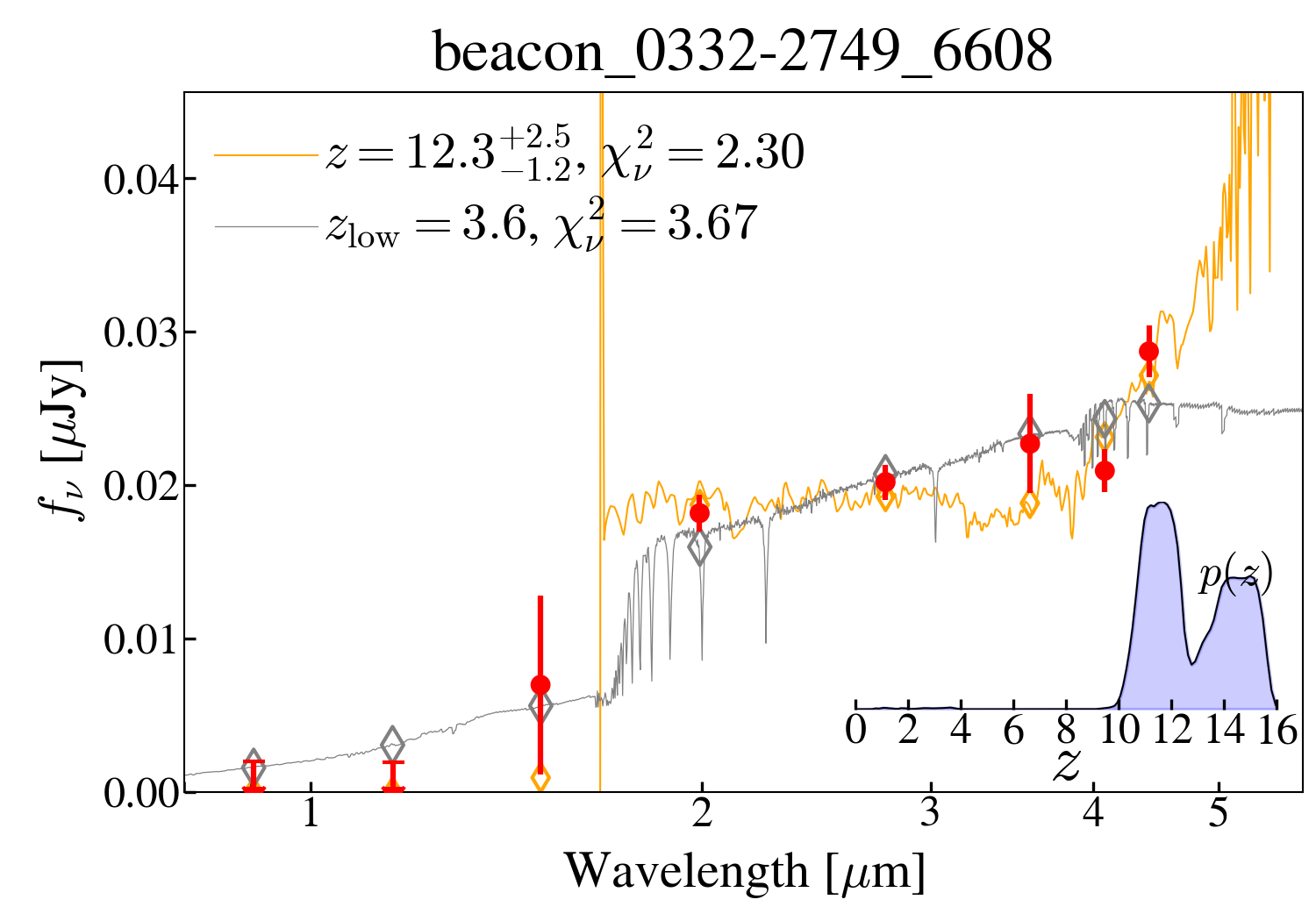}
	\caption{Postage stamps and SEDs of example high-$z$ galaxy candidates. The best-fit flux models at high redshift (orange lines) and at the secondary low redshift (z$_{\rm low}$, gray line) are shown. Photometric redshift probability distributions derived from {\tt EAzY} are shown in the inset. The y-axis is set in a linear scale. 
    }
\label{fig:cand}
\end{figure*}

The cutout images and SEDs of example $z\simgt8$ candidates are shown in Fig.~\ref{fig:cand}. At the redshift range of $z<10$, we highlight two UV luminous galaxies. These examples showcase a variety of spectral features: ID336 is best fitted with an extreme emission line template, characterized by a blue rest-frame UV slope, and flux excess in the F444W band. ID2768 is located in the field beacon\_1420+5252, where several medium bands are also taken (F140M, F182M, F410M, F430M, F480M). Red medium-band filters sample the wavelength around the Balmer break and thus effectively discriminate other possible solutions, e.g., line excess due to strong \hb+\oiii\ emission lines, as seen in ID336.

We have identified 11 $z\simgt10$ galaxies of moderate brightness, $-M_{\rm UV}\sim18.5$--$20.5$. Although the photometric flux errors are relatively large, leading to larger $z$ uncertainties, all of these candidates are secured with a clear dropout in the F090W and F115W bands. At $z>9$, \hb+\oiii\ lines are redshifted to $>5\,\mu$m and thus a relatively flat slope is expected at $\sim2$--5\,$\mu$m. In Fig.~\ref{fig:cand}, we show two example sources from the redshift. ID13052 has a secondary solution at $z\sim2.4$. The existing photometry, however, the low-$z$ solution at $\sim2\,\mu$m discriminates. ID6608 is one of the highest-redshift candidates among our final sample. It has a relatively wide redshift probability distribution at $z>10$, which is partially attributed to the fact that the spectrum is rather featureless for the observed wavelength range. However, the redshift probability at $z<10$, with a secondary solution at $z\sim3.6$, is low ($p(z<10)=0.004$), making it a likely high-$z$ candidate.  

Also noted is that most of the selected candidates have extended morphologies. While the multi-band NIRCam filters up to $\sim5\,\mu$m effectively eliminates the contamination of fore-ground low-mass stars (see Sec.~\ref{sec:anc}), the morphological information (e.g., size, elongation, light concentration) of the color-selected source can further secure the final high-$z$ sample (Zhang, in prep.).

\subsection{Number Densities of UV-selected Galaxies at $z>7$}\label{sec:nd}
Using the selected candidates, we aim to estimate the number density of galaxies at $z>7$. An unusual aspect of our survey is that each field has different depths and filters. This results in completeness and selection functions that vary from field to field, requiring a careful effective volume estimate in each field.

For the effective volume estimate, we use an adaptation of the \texttt{GLACiAR2} completeness simulation code \citep{nicha22,leethochawalit23}. Briefly, in each field, we inject 1200 galaxies into each magnitude-redshift bin. The bins run from $z=6$ to $18$ and from $M_{UV}=-18.5$ to $-24$\,mag with an increment of 0.5 in both directions. The injected galaxies are assumed to have a S\'ersic profile of $n=1$ with the sizes that follow the $M_{UV}$-size relation as a function of redshift in \citet{morishita24}. The spectral shapes of the galaxies are randomly pulled from the JAGUAR \citep{Williams2018} catalogs of the same redshift bin up to $z=12$. Since there are not many spectra at $z>12$ in the JAGUAR catalog, we combine all $z>12$ JAGUAR spectra into one spectral pool. The images of the injected galaxies are convolved with the PSF from the F444W band and injected into the PSF-matched images. The detection, photometric extraction, and sample selection of the injected galaxies are done in the same manner as the real sources. 

We then calculate the effective volume of each field as in \citet{leethochawalit23}:
\begin{equation}
    V_\textrm{eff}(M_\textrm{recov}) = \int\frac{dV}{dz}P(M_\textrm{recov},z) dz.
    \label{eq:veff_mrecov}
\end{equation}
$P(M_\textrm{recov},z)$ is the number of simulated galaxies at redshift $z$ that are recovered to have UV magnitude in the bin $M_\textrm{recov}$ (regardless of their intrinsic UV magnitude), divided by the number of injected galaxies at redshift $z$ with intrinsic UV magnitude equal to $M_\textrm{recov}$. 
For reference, at the brightest magnitude bins (i.e. $\sim100$\,\% completeness), the effective volumes for a single NIRCam pointing are approximately $2\times10^4$\,Mpc$^3$ for all redshift windows. 
Once effective volumes are calculated, we calculate the number density at each absolute UV magnitude bin by combining all fields. 

\begin{figure*}
\centering
	\includegraphics[width=0.46\textwidth]{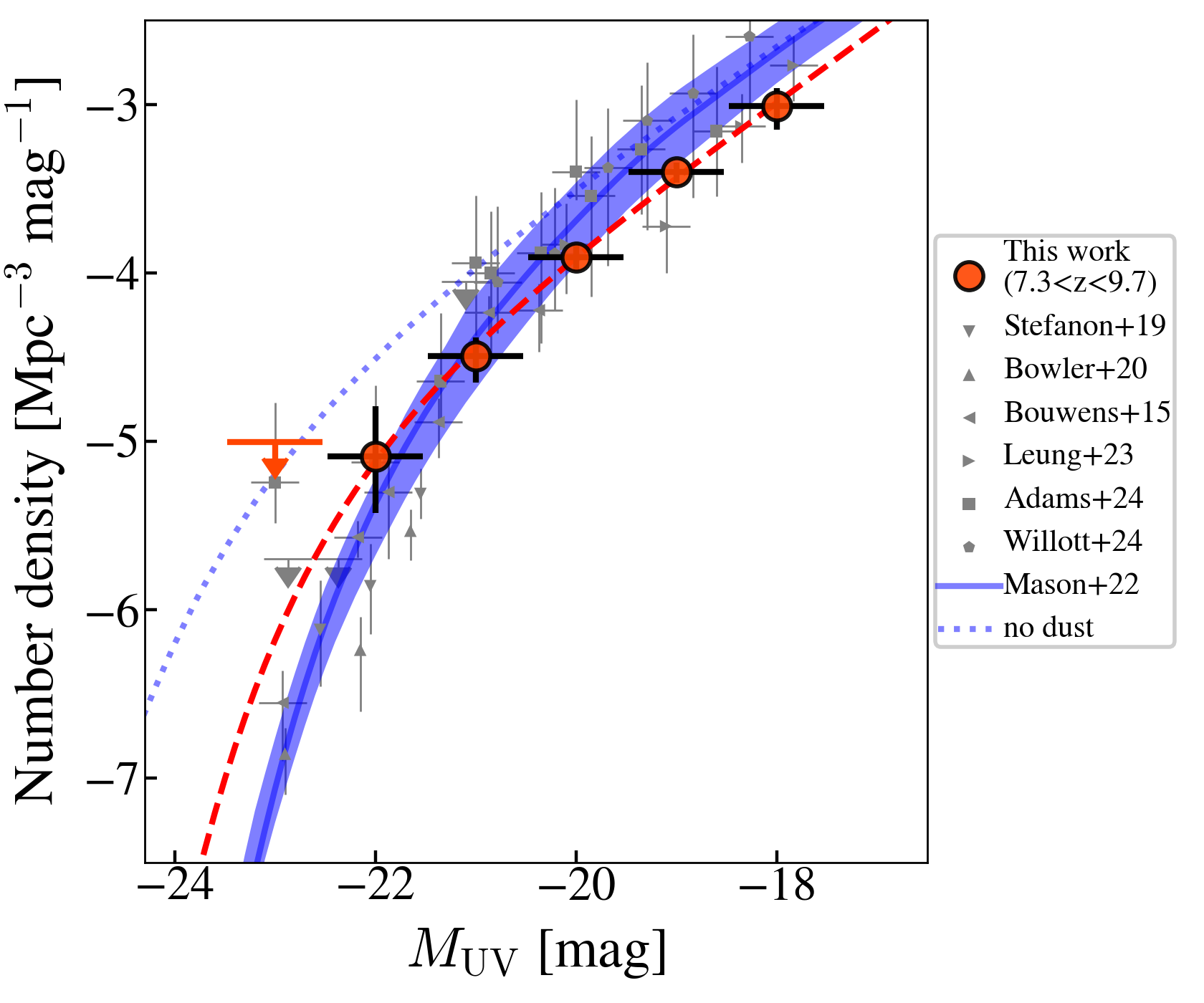}
	\includegraphics[width=0.46\textwidth]{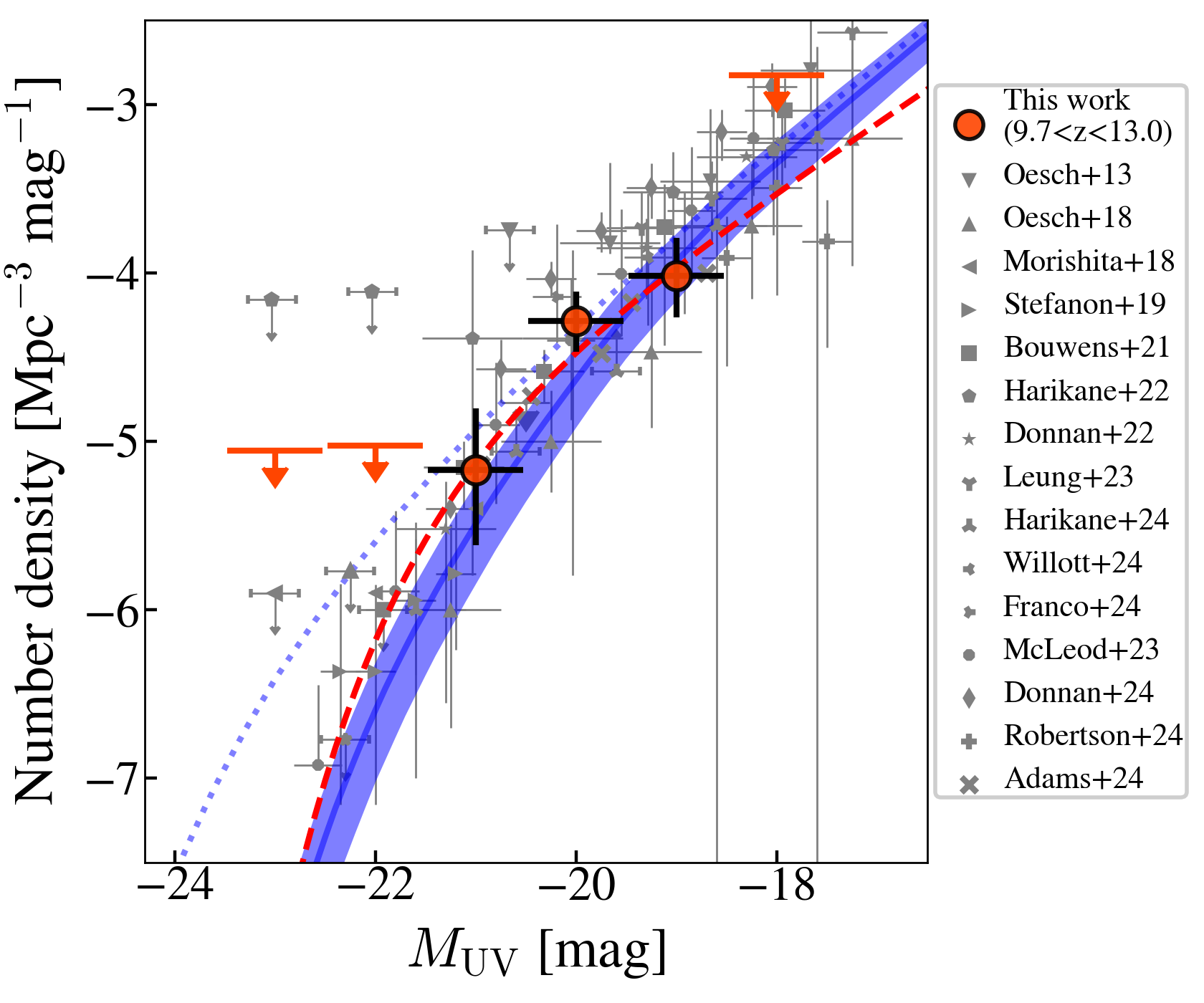}
	\includegraphics[width=0.46\textwidth]{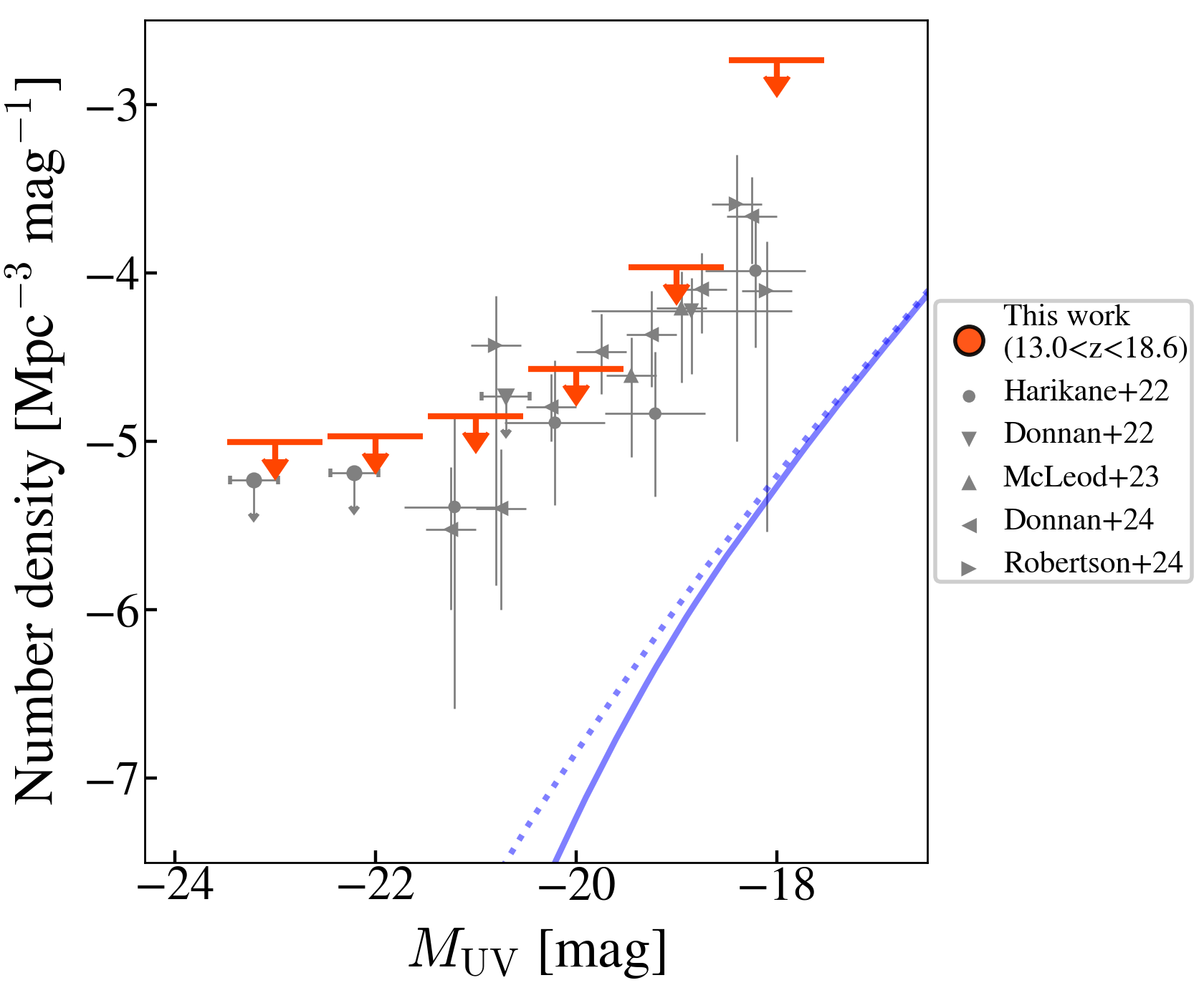}
	\caption{
 Number densities estimated for the Beacon DR1 sources (red circles and upper limits), at $z\sim8.5$ (left top), $z\sim11$ (right top), and $z\sim15$ (bottom). For magnitude bins without any sources, we show 2-$\sigma$ upper limits. The Schechter fit to our number density estimates is shown (red dashed lines). Number density estimates in the literature are shown \citep[gray symbols;][]{oesch13,bouwens15,bouwens21,morishita18b,stefanon19,Donnan2022,harikane22b,harikane23,mcleod24, leung23,willott24,Rojas-Ruiz24,franco24,adams24}.
 The \citet{mason22} LFs (blue solid lines for the dust model, dashed lines for the no-dust model) of the corresponding redshift are shown. The redshift variation of the dust model is shown by a hatched region.
    }
\label{fig:LF}
\end{figure*}

The estimated galaxy number densities are shown in Fig.~\ref{fig:LF}. Our estimates are overall consistent with studies in the literature at $7<z<13$ \citep{harikane22b,Donnan2022,Bouwens2022b,casey23,franco24,willott24,adams24}. Our estimates are also consistent with the theoretical prediction of \citet{mason22}. The total effective volumes and the calculated number densities are listed in Tab.~\ref{tab:nd}.

To assess our results in more detail, we fit the number density estimates with the \citet{schechter76} function:
\begin{equation}
    \begin{aligned}
            \phi(M_{\rm UV}) &= {{\ln 10} \over{2.5}} \phi^* \times 10^{0.4(\alpha+1)(M_{\rm UV}-M^*)}\\
            &\times {\rm exp}\big[-10^{-0.4 (M_{\rm UV}-M^*)}\big]
    \end{aligned}
\end{equation}
using a Markov Chain Monte Carlo (MCMC) method \citep[e.g.,][]{morishita18b}. The derived parameters for the $7.3<z<9.7$ number density are $\alpha = -2.07_{-0.18}^{+0.23}, \log\phi^* = -4.66_{-0.44}^{+0.53}$, and $M^* = -21.97_{-0.66}^{+0.73}$. For the $9.7<z<13$ range, we fix the slope to $\alpha = -2$ and find $\log\phi^* = -4.57_{-0.96}^{+2.10}$ and $M^* = -21.01_{-1.45}^{+1.61}$.

We estimate that we will find $\sim650$ galaxies at $7.3<z<9.7$ and $\sim100$ at $9.7<z<13$ in the full BEACON dataset, by using our LF estimates and simply scaling the survey volume while assuming 80\% completeness down to the 10$\sigma$ limiting magnitude. Our LF predicts similar numbers of galaxies to those of the constant star formation efficiency model by \citet{mason22} as in Fig.~\ref{fig:ND}.

Despite covering a considerably large total area ($\sim\area$\,arcmin$^2$), there are no $z>13$ galaxy candidates selected in our final samples. We place upper limits for the number densities (Fig.~\ref{fig:LF}). The upper limits derived are still consistent with previous studies in legacy fields \citep{harikane22b,donnan24,mcleod24}, except for \citet{robertson24} at the bright end ($M_{\rm UV}\sim-21$) which contains the luminous $z=14$ galaxy GS-z14-0. The absence of luminous $z>13$ galaxies in our fields may suggest a significant impact from cosmic variance affecting the statistics in those legacy fields. As we see in Sec.~\ref{sec:ov}, a large fraction of $z_{\rm spec}>10$ sources are located just in $two$ fields. In addition, exceptionally luminous sources in the literature (GN-z11, GHZ2/GLASS-z12, GS-z14-0) are found in fields of relatively high cosmic variance ($\sigma_\mathrm{cv}>0.4$; Fig.~\ref{fig:cv}). The complete data set of BEACON will reach a $5\times$ larger volume ($\sim1.5\times10^6$\,Mpc$^3$) and facilitate us to explore this further.

\subsection{Ancillary Science}\label{sec:anc}
\begin{figure*}
\centering
	\includegraphics[width=0.49\textwidth]{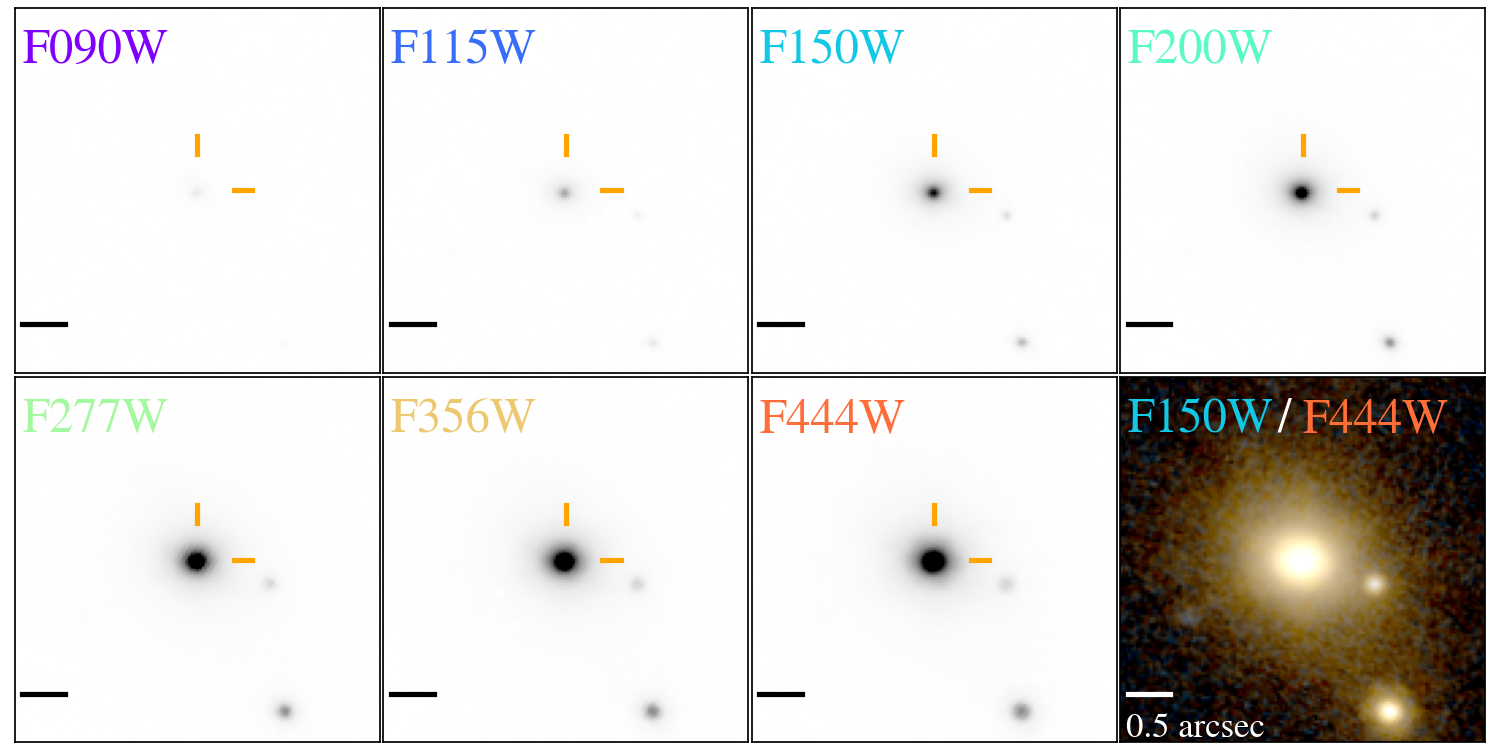}
	\includegraphics[width=0.49\textwidth]{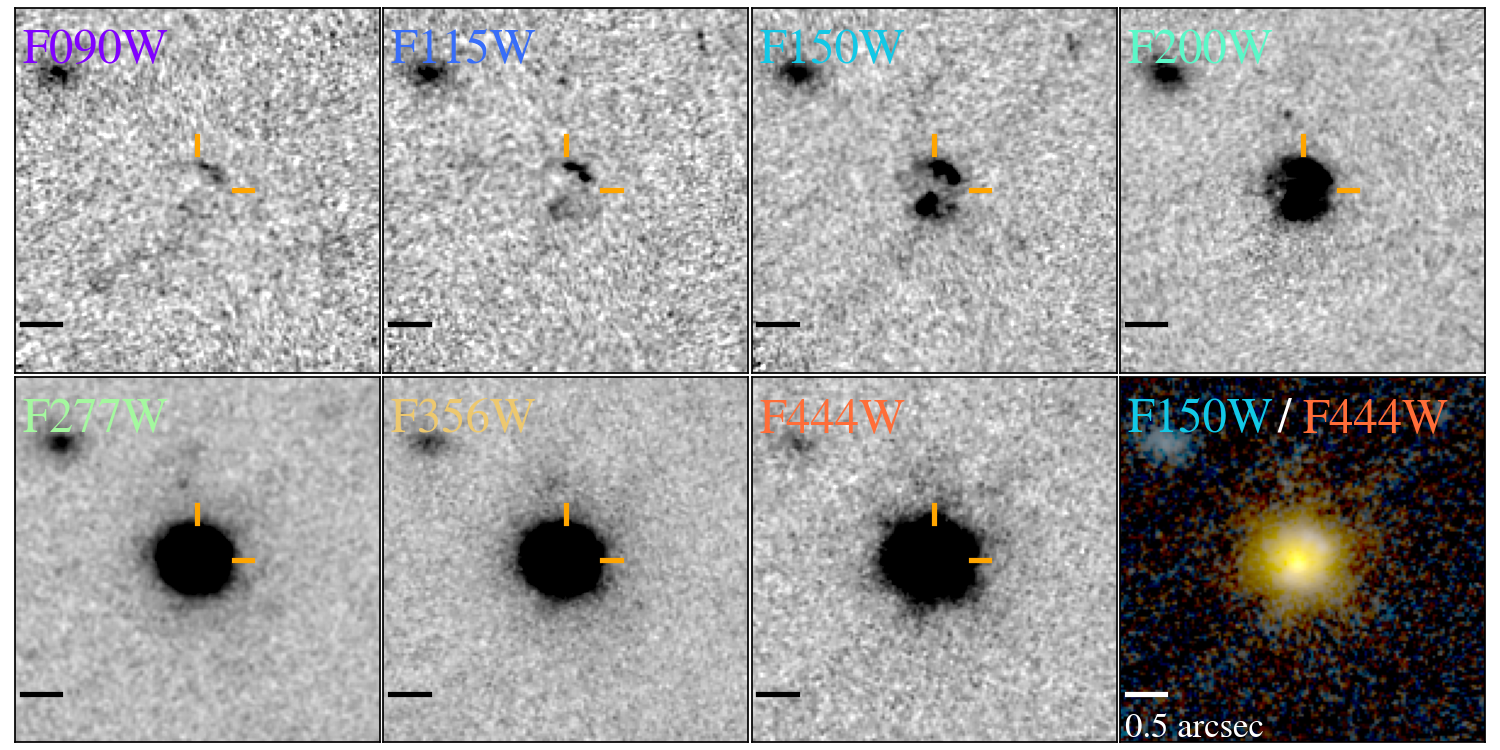}
	\includegraphics[width=0.49\textwidth]{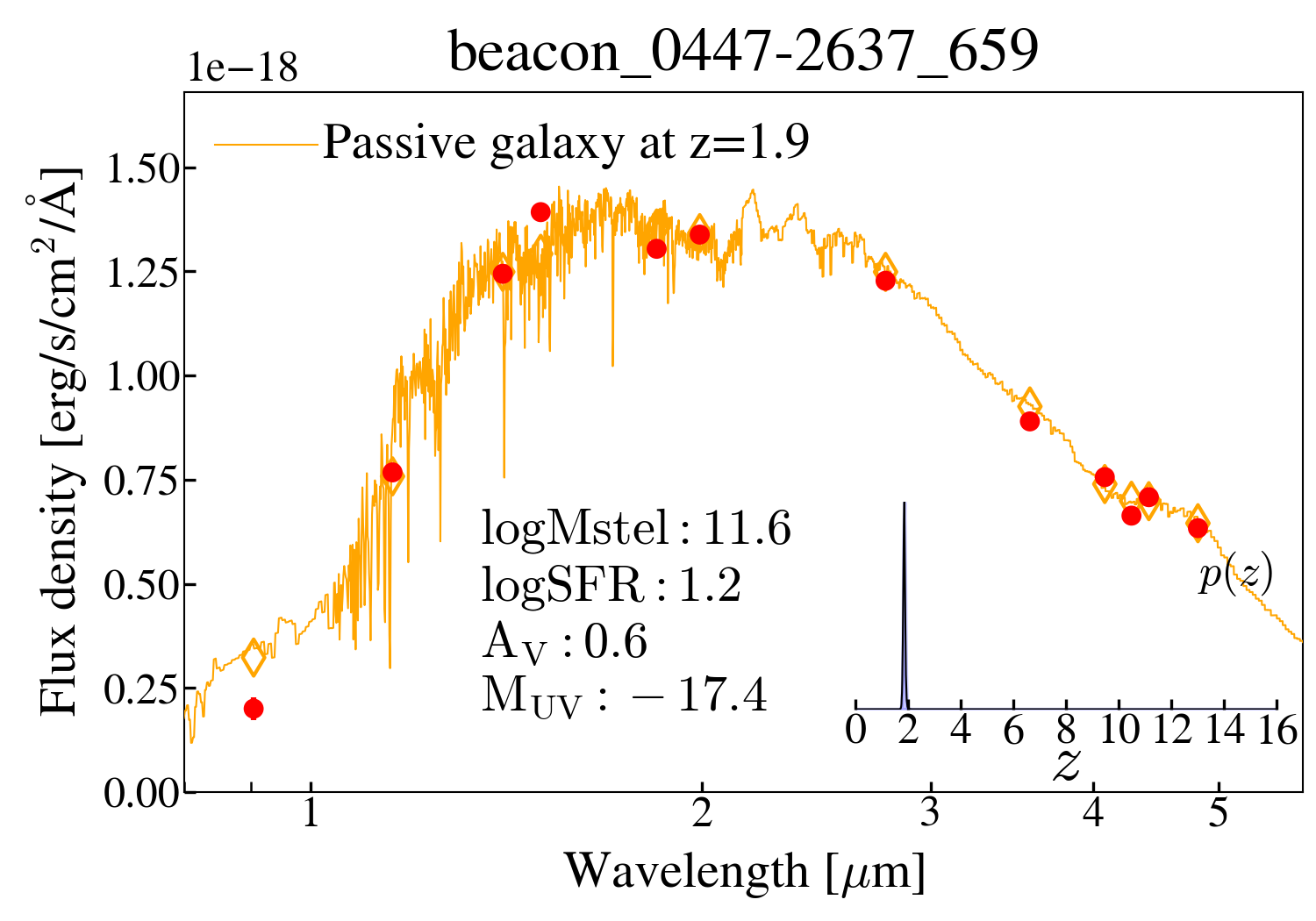}
	\includegraphics[width=0.49\textwidth]{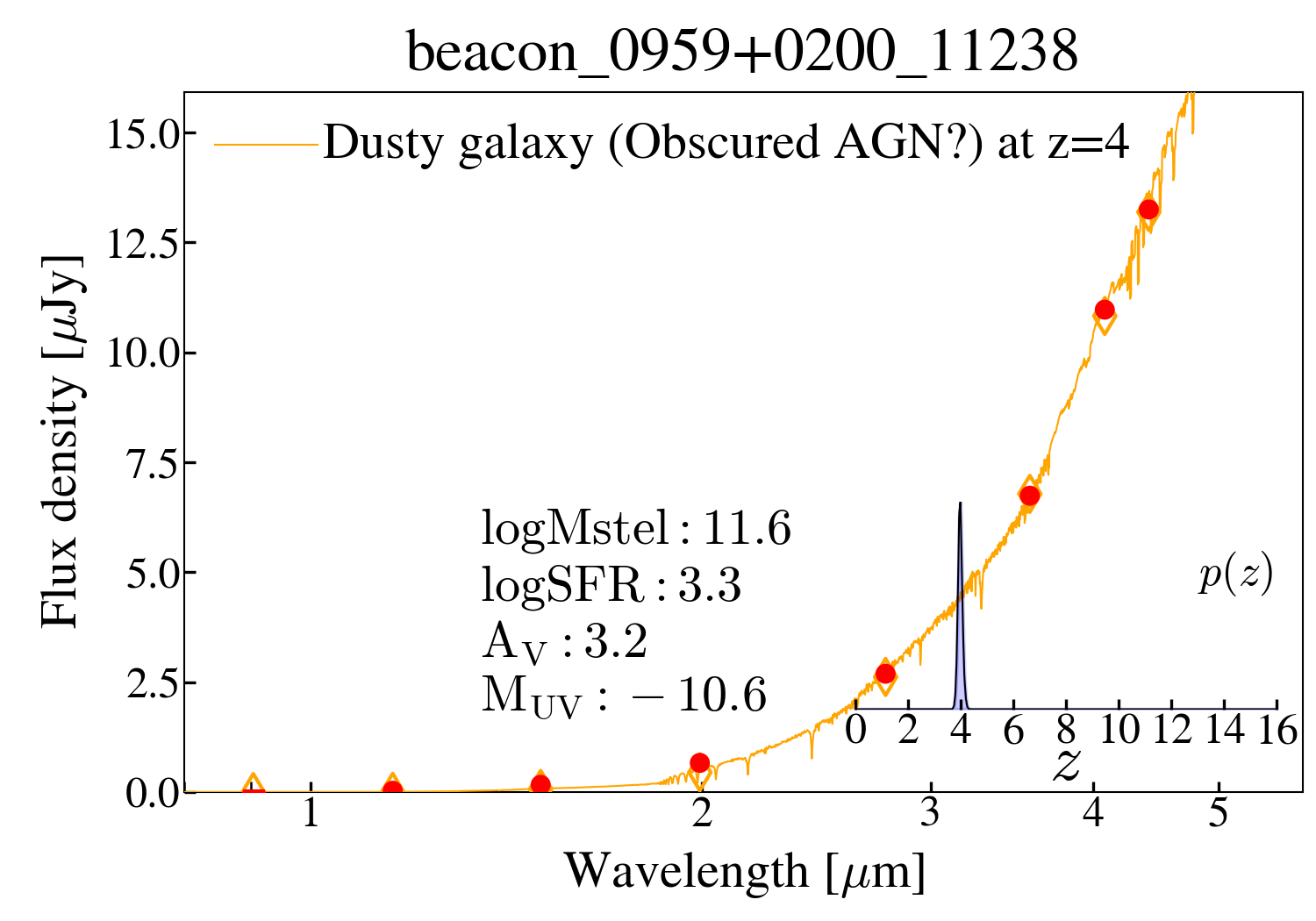}
	\includegraphics[width=0.5\textwidth]{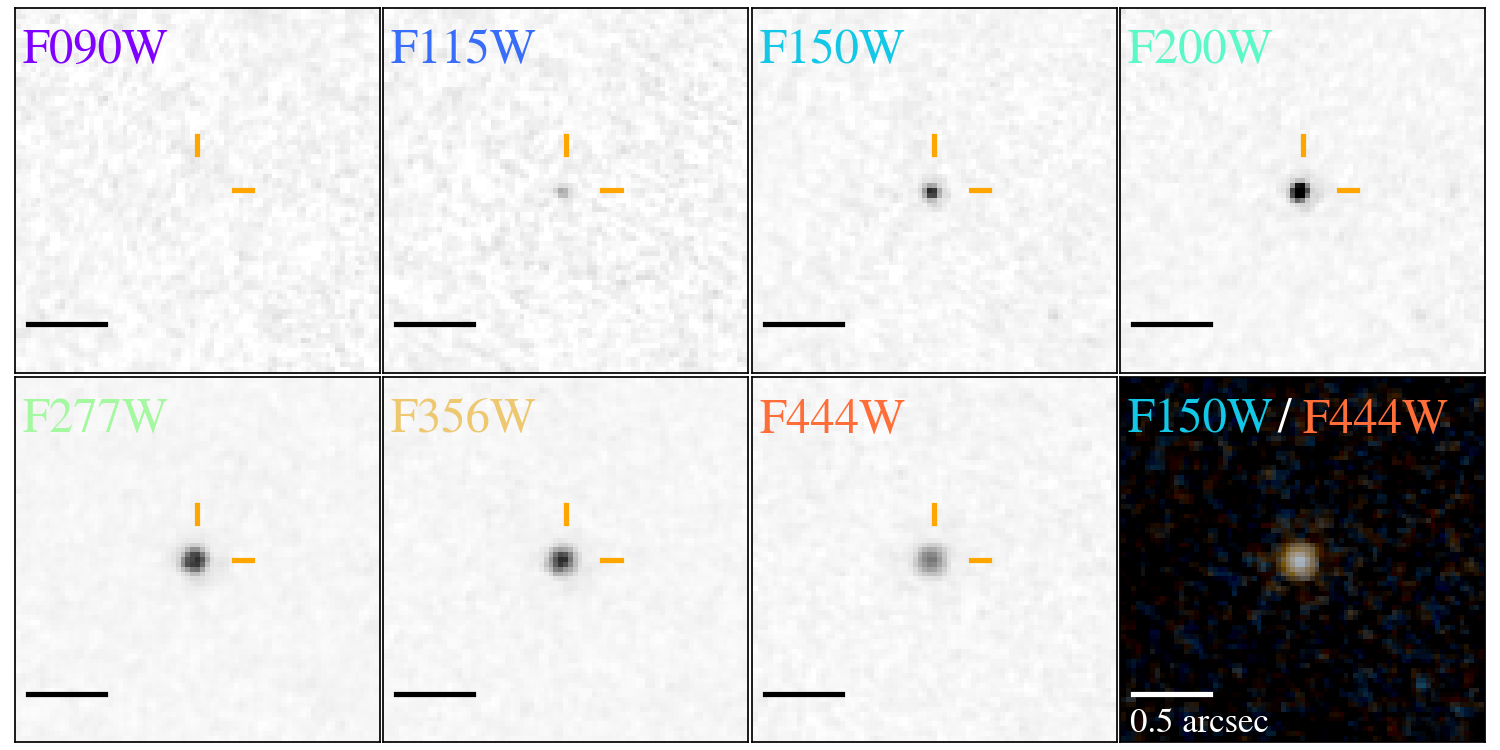}
	\includegraphics[width=0.5\textwidth]{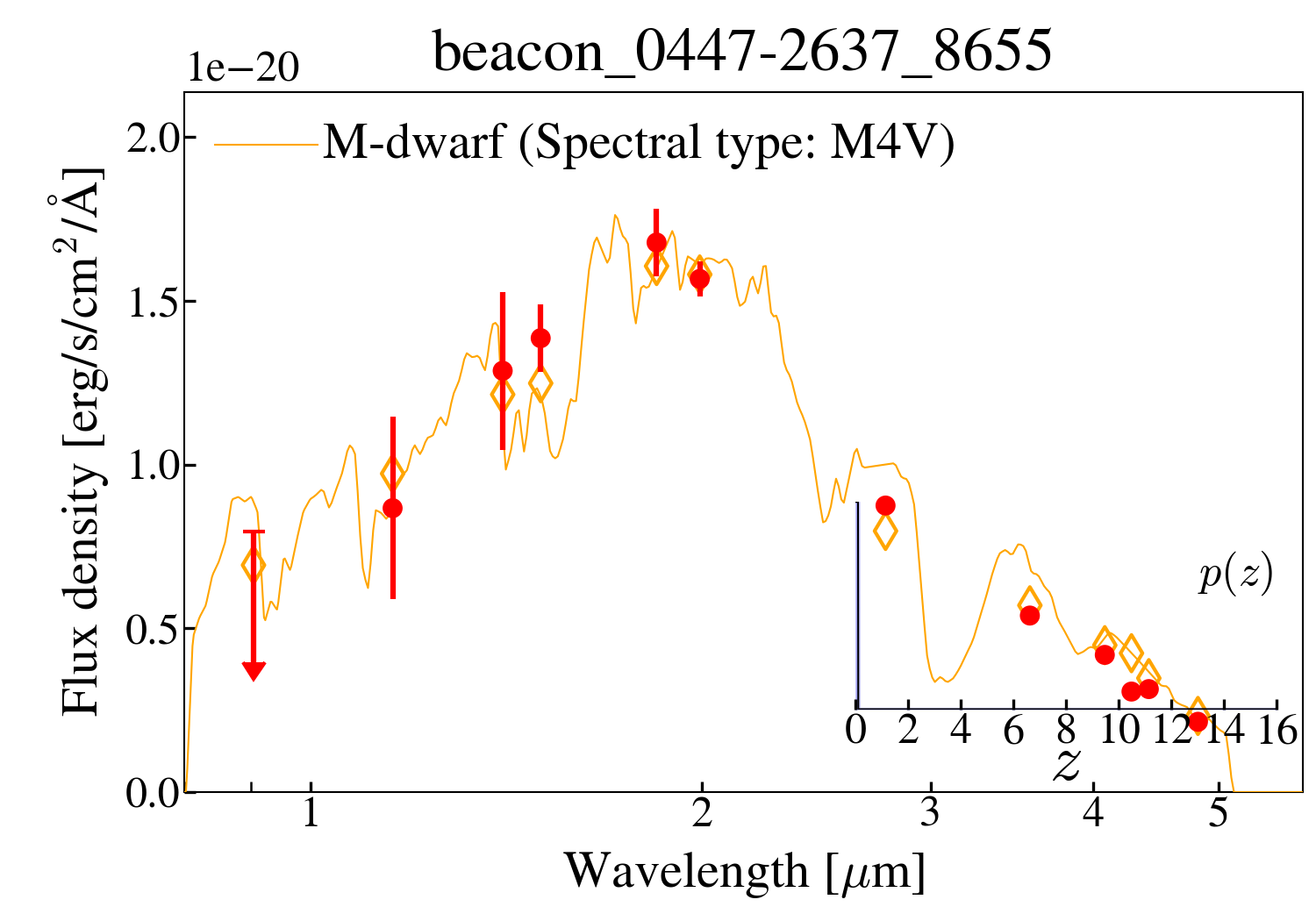}
	\caption{Examples of low-$z$ sources from BEACON. The figure format is the same as in Fig.~\ref{fig:cand}. 
 $Left$: Massive ($\logm\sim11.6$), passively evolving galaxy at $z\sim1.9$. 
 $Right$: Dusty galaxy at $z\sim4$. The galaxy is characterized with a red spectral shape ($A_V\sim3.2$). It has a compact point source (possibly obscured AGN) dominating at $>2\,\mu$m, whereas extended structures are seen in bluer wavelengths.
 $Bottom$: faint ($m_{\rm F200W}=25.6$), M4V dwarf selected from point-source analysis and spectral template fitting. The best-fit spectral template (M4V) from the {\tt SPEX} library (yellow line) is shown.
}
\label{fig:anc}
\end{figure*}

Our pure-parallel survey will provide a unique and extensive dataset for legacy science at lower redshifts. In particular, our filter coverage can secure key spectral features, such as the Balmer break of $z\sim2$ galaxies (Fig.~\ref{fig:filter}) and the $3.3\,\mu$m PAH emission line of $z\sim0.3$ galaxies seen as a flux excess in F410M \citep[e.g.,][]{vulcani23}. In Fig.~\ref{fig:anc}, we show two galaxy examples selected from the DR1 fields: a passively evolving galaxy at $z\sim1.9$ and a dusty galaxy at $z\sim4$. Also, foreground low-mass stars (brown dwarfs) are often found as contaminants in high-$z$ galaxy color selection, for its similar spectral features i.e. the color break at observed $\sim1$\,$\mu$m \citep[][also \citealt{greene23} for spectroscopic confirmation of faint brown dwarf stars]{morishita20,ishikawa22}. The NIRCam data from BEACON securely discriminate those stars from galaxy populations by adding longer wavelength data points at $>2$\,$\mu$m. The secure selection of faint stars enables us to study, e.g., their spatial distribution in the Milky Way halo \citep[e.g.,][]{ryan11,holwerda14b}. In Fig.~\ref{fig:anc}, we show one M-dwarf candidate, selected from spectral template fitting using the {\tt SPEX} library \citep{rayner03} and additional morphological analysis for point source selection.

Lastly, we expect some fraction of our fields to overlap with the existing survey fields where previous NIRCam fields are available. This is because long-exposure spectroscopic observations tend to target sources found within those popular legacy fields. In the DR1 fields, seven are found within the legacy fields (Fig.~\ref{fig:sky}). Those overlapping fields are ideal for the detection of transient events and variability studies \cite[e.g.,][]{kokubo24,zhang24}.

\section{Summary}\label{sec:sum}
In this paper, we introduced BEACON, a new pure-parallel imaging survey scheduled in JWST Cycle 2. Using the first \nfld\ fields, we established our data reduction processes and performed an initial high-$z$ galaxy search. Through our careful color selection utilizing 6--8+ NIRCam filters, we identified \nsrc\ galaxy candidates at $7<z<13$. The galaxy number densities estimated in the fields are overall consistent with the previous estimates in the literature. Although our exploration reached a considerably large area ($\sim$\,\area\,arcmin$^2$), we found zero $z>13$ galaxies. However, the upper limits of the $z>13$ galaxy number density are consistent with previous studies in the literature. Further exploration utilizing the full BEACON dataset will confidently determine the contribution of cosmic variance to the number densities of bright $z>13$ galaxies detected so far with JWST.

The BEACON dataset also allows for a range of lower-$z$ science cases, such as the search for massive galaxies at cosmic noon and foreground low-mass stars. We plan to process all imaging data and publish the resulting products, including photometric flux redshift catalogs for each survey field. We will also make available the effective volume estimate from our completeness simulation. We will publish the data products of the DR1 fields on a dedicated website. Catalog data models are presented in Tab.~\ref{tab:col}. The full data release will be made when the program is complete.

\begin{deluxetable}{cccc}
\tabletypesize{\footnotesize}
\tabcolsep=12pt
\tablecolumns{4}
\tablewidth{0pt} 
\tablecaption{Number density of galaxies at $z>8$}
\tablehead{\colhead{$M_{\rm UV}$} & \colhead{Volume$^{\rm a}$} & \colhead{$N_{\rm obj}$} & \colhead{Number density$^{\rm b}$}
\vspace{-0.3cm}\\
\colhead{(mag)} & \colhead{($10^3$\,Mpc$^{3}$)} & \colhead{} & \colhead{($\log$\,Mpc$^{-3}$ mag$^{-1}$)}
}
\startdata
\cutinhead{F090W-dropout}
$-23.0$ & 366.9 & 0 & $<-5.30$\\
$-22.0$ & 366.5 & 3 & $-5.09_{-0.29}^{+0.34}$\\
$-21.0$ & 344.6 & 11 & $-4.50_{-0.11}^{+0.16}$\\
$-20.0$ & 282.0 & 33 & $-3.93_{-0.07}^{+0.08}$\\
$-19.0$ & 143.0 & 54 & $-3.42_{-0.06}^{+0.06}$\\
$-18.0$ & 13.3 & 13 & $-3.01_{-0.11}^{+0.14}$\\
$-17.0$ & 0.1 & 0 & $<-1.74$\\
\cutinhead{F115W-dropout}
$-23.0$ & 394.4 & 0 & $<-5.33$\\
$-22.0$ & 365.1 & 0 & $<-5.30$\\
$-21.0$ & 276.2 & 2 & $-5.14_{-0.36}^{+0.45}$\\
$-20.0$ & 152.2 & 8 & $-4.28_{-0.17}^{+0.18}$\\
$-19.0$ & 52.1 & 5 & $-4.02_{-0.22}^{+0.24}$\\
$-18.0$ & 2.4 & 0 & $<-3.13$\\
$-17.0$ & 0.1 & 0 & $<-1.64$\\
\cutinhead{F150W-dropout}
$-23.0$ & 355.7 & 0 & $<-5.29$\\
$-22.0$ & 328.1 & 0 & $<-5.25$\\
$-21.0$ & 249.3 & 0 & $<-5.14$\\
$-20.0$ & 133.3 & 0 & $<-4.86$\\
$-19.0$ & 33.9 & 0 & $<-4.27$\\
$-18.0$ & 2.0 & 0 & $<-3.04$\\
$-17.0$ & 0.1 & 0 & $<-1.89$\\
\enddata
\label{tab:nd}
\tablenotetext{\rm {\bf Notes.}
}
{
\\
$\rm ^a$ Effective volume calculated by the completeness simulation.\\
$\rm ^b$ 1~$\sigma$ uncertainty calculated for small number statistics as in \citet{gehrels86}.
}
\end{deluxetable}


\section*{Acknowledgements}
We thank the program coordinators of our JWST program, Shelly Meyett and Blair Porterfield, and the instrument scientist, Anton Koekemoer, for their invaluable assistance in configuring our observations. In addition to being one of the largest JWST programs, the complexity of the pure-parallel mode introduced a series of unique challenges and issues, each of which benefited greatly from their expertise and perspectives. We also thank the Cycle~1 and 2 pure-parallel program teams (PASSAGE, PANORAMIC, and PID~3383 (PI Glazebrook)) for their pioneering efforts, which laid the essential foundation and facilitated the operation of pure-parallel observations. We thank Xuejian Shen (MIT) and the THESAN team for sharing dark-matter density maps and related physical property catalogs from their simulations. TM and YZ thank George Helou for generously supporting our data analysis at IPAC. 
Support for program 3990 was provided by NASA through the Space Telescope Science Institute, which is operated by the Association of Universities for Research in Astronomy, Inc., under NASA contract NAS 5-03127. 
All of the data presented in this paper were obtained from the Mikulski Archive for Space Telescopes (MAST) at the Space Telescope Science Institute. The specific observations analyzed can be accessed via \dataset[10.17909/q8cd-2q22]{https://doi.org/10.17909/q8cd-2q22}.
CAM, KCK and VG acknowledge support from the Carlsberg Foundation under grant CF22-1322. The Cosmic Dawn Center (DAWN) is funded by the Danish National Research Foundation under grant DNRF140.
BV acknowledges support from the European Union – NextGenerationEU RFF M4C2 1.1 PRIN 2022 project 2022ZSL4BL INSIGHT. MB acknowledges support from the ERC Grant FIRSTLIGHT and Slovenian national research agency ARIS through grants N1-0238 and P1-0188.

{
{\it Software:} 
AstroML \citep{astroML}, Astropy \citep{astropy13,astropy18,astropy22}, bbpn \citep{bbpn}, EAzY \citep{brammer08}, EMCEE \citep{foreman13}, numpy \citep{numpy}.
}


\startlongtable
\begin{deluxetable*}{lcccccc}
\tabletypesize{\footnotesize}
\tablecolumns{6}
\tablewidth{0pt}
\tablecaption{Selected high-$z$ galaxy candidates.
}
\tablehead{
\colhead{ID} & \colhead{R.A.} & \colhead{Decl.} & \colhead{$m_{\rm F150W}$} & \colhead{$m_{\rm F444W}$} & \colhead{$z$} & \colhead{$M_{\rm UV}$}
}
\startdata
beacon\_0217-0509-1447 & $34.225235$ & $-5.182756$ & 28.1 & 27.5 & $7.04_{-0.18}^{+0.20}$ & $-19.01_{-0.02}^{+0.01}$ \\
beacon\_0217-0508-9736 & $34.293415$ & $-5.115285$ & 28.4 & 28.7 & $7.04_{-0.18}^{+0.37}$ & $-18.74_{-0.01}^{+0.01}$ \\
beacon\_1420+5252-7386 & $215.037430$ & $52.877472$ & 28.2 & 27.4 & $7.08_{-0.08}^{+0.09}$ & $-18.94_{-0.01}^{+0.01}$ \\
beacon\_1420+5252-12023 & $215.039520$ & $52.891056$ & 28.1 & 27.8 & $7.13_{-0.13}^{+0.11}$ & $-19.06_{-0.01}^{+0.01}$ \\
beacon\_1420+5252-7298 & $215.038239$ & $52.877132$ & 27.7 & 27.4 & $7.13_{-0.13}^{+0.11}$ & $-19.20_{-0.01}^{+0.01}$ \\
beacon\_1420+5252-9579 & $215.037079$ & $52.892605$ & 27.2 & 26.7 & $7.14_{-0.11}^{+0.10}$ & $-19.96_{-0.01}^{+0.01}$ \\
beacon\_1420+5252-3062 & $215.040863$ & $52.863846$ & 27.8 & 27.5 & $7.14_{-0.15}^{+0.12}$ & $-19.31_{-0.01}^{+0.01}$ \\
beacon\_1420+5252-10296 & $215.125000$ & $52.874062$ & 26.2 & 25.8 & $7.16_{-0.12}^{+0.11}$ & $-20.92_{-0.01}^{+0.01}$ \\
beacon\_0217-0504-6307 & $34.273205$ & $-5.037470$ & 28.2 & 27.1 & $7.17_{-0.17}^{+0.19}$ & $-18.82_{-0.02}^{+0.02}$ \\
beacon\_0217-0509-4631 & $34.241405$ & $-5.150918$ & 28.6 & 28.6 & $7.18_{-0.36}^{+0.45}$ & $-18.61_{-0.06}^{+0.05}$ \\
beacon\_0959+0200-14013 & $149.833344$ & $2.036179$ & 27.5 & 27.4 & $7.18_{-0.18}^{+0.23}$ & $-19.50_{-0.01}^{+0.01}$ \\
beacon\_1420+5252-12575 & $215.025375$ & $52.891273$ & 27.2 & 26.1 & $7.18_{-0.07}^{+0.05}$ & $-19.91_{-0.01}^{+0.01}$ \\
beacon\_0217-0504-2657 & $34.272232$ & $-5.045108$ & 28.9 & 27.6 & $7.19_{-0.12}^{+0.24}$ & $-18.14_{-0.01}^{+0.01}$ \\
beacon\_1420+5252-1569 & $214.994904$ & $52.866634$ & 27.6 & 27.3 & $7.20_{-0.19}^{+0.11}$ & $-19.54_{-0.01}^{+0.01}$ \\
beacon\_1138+5748-6058 & $174.370514$ & $57.791782$ & 26.8 & 26.0 & $7.26_{-0.11}^{+0.11}$ & $-20.25_{-0.01}^{+0.01}$ \\
beacon\_0959+0200-92 & $149.819672$ & $1.965411$ & 27.2 & 26.8 & $7.29_{-0.18}^{+0.16}$ & $-19.98_{-0.01}^{+0.01}$ \\
beacon\_0860+3857-14167 & $134.982468$ & $38.985676$ & 27.0 & 26.1 & $7.29_{-0.13}^{+1.05}$ & $-20.02_{-0.01}^{+0.16}$ \\
beacon\_0217-0508-8517 & $34.283566$ & $-5.124747$ & 29.1 & 29.0 & $7.32_{-0.13}^{+0.30}$ & $-17.94_{-0.01}^{+0.01}$ \\
beacon\_0332-2745-7700 & $53.071823$ & $-27.782648$ & 27.8 & 27.4 & $7.34_{-0.15}^{+0.49}$ & $-19.44_{-0.01}^{+0.04}$ \\
beacon\_0959+0200-1159 & $149.840607$ & $1.964149$ & 27.9 & 27.3 & $7.34_{-0.17}^{+0.14}$ & $-19.48_{-0.01}^{+0.01}$ \\
beacon\_0014-3025-10125 & $3.625879$ & $-30.439075$ & -- & 27.9 & $7.35_{-0.07}^{+0.60}$ & $-18.52_{-0.01}^{+0.08}$ \\
beacon\_0332-2745-8900 & $53.053425$ & $-27.765003$ & 27.8 & 26.8 & $7.35_{-0.10}^{+0.08}$ & $-19.19_{-0.01}^{+0.01}$ \\
beacon\_1138+5748-3075 & $174.362335$ & $57.780087$ & 27.3 & 26.9 & $7.37_{-0.47}^{+0.54}$ & $-19.78_{-0.06}^{+0.08}$ \\
beacon\_1010+2701-3159 & $152.407913$ & $27.005966$ & 26.9 & 26.9 & $7.38_{-0.16}^{+0.29}$ & $-20.43_{-0.01}^{+0.01}$ \\
beacon\_0332-2745-15094 & $53.046597$ & $-27.738073$ & 27.3 & 27.5 & $7.39_{-0.15}^{+0.20}$ & $-19.79_{-0.01}^{+0.01}$ \\
beacon\_0332-2749-5025 & $53.086330$ & $-27.823927$ & 27.4 & 26.6 & $7.39_{-0.10}^{+0.35}$ & $-19.69_{-0.01}^{+0.01}$ \\
beacon\_0959+0200-1349 & $149.829239$ & $1.969253$ & 28.2 & 28.5 & $7.39_{-0.38}^{+0.43}$ & $-18.77_{-0.04}^{+0.03}$ \\
beacon\_0332-2745-13954 & $53.086193$ & $-27.773207$ & 28.6 & 28.4 & $7.44_{-0.36}^{+0.42}$ & $-18.70_{-0.04}^{+0.04}$ \\
beacon\_0217-0504-17278 & $34.329090$ & $-5.092696$ & 28.7 & 28.5 & $7.45_{-0.42}^{+0.53}$ & $-18.51_{-0.06}^{+0.05}$ \\
beacon\_0217-0504-6678 & $34.284660$ & $-5.053731$ & 29.0 & 29.0 & $7.45_{-0.35}^{+0.42}$ & $-18.18_{-0.03}^{+0.04}$ \\
beacon\_1420+5252-3630 & $215.084747$ & $52.858551$ & 27.7 & 27.6 & $7.45_{-0.08}^{+0.07}$ & $-19.72_{-0.01}^{+0.01}$ \\
beacon\_0217-0508-1040 & $34.320229$ & $-5.164967$ & 28.1 & 28.7 & $7.48_{-0.36}^{+0.45}$ & $-19.14_{-0.04}^{+0.03}$ \\
beacon\_0217-0504-10666 & $34.311810$ & $-5.084851$ & 28.0 & 26.8 & $7.49_{-0.23}^{+0.92}$ & $-19.31_{-0.03}^{+0.08}$ \\
beacon\_1420+5252-11406 & $215.110046$ & $52.872971$ & 28.5 & 28.6 & $7.51_{-0.47}^{+0.55}$ & $-18.77_{-0.04}^{+0.05}$ \\
beacon\_0217-0504-18177 & $34.323090$ & $-5.081106$ & 27.7 & 26.9 & $7.51_{-0.35}^{+0.73}$ & $-19.49_{-0.04}^{+0.07}$ \\
beacon\_0959+0200-13202 & $149.845184$ & $2.029492$ & 27.6 & 27.4 & $7.53_{-0.42}^{+0.43}$ & $-19.61_{-0.04}^{+0.05}$ \\
beacon\_0332-2745-10227 & $53.062523$ & $-27.768053$ & 28.2 & 27.5 & $7.54_{-0.50}^{+0.38}$ & $-19.11_{-0.07}^{+0.05}$ \\
beacon\_0332-2745-8698 & $53.053013$ & $-27.765186$ & 27.9 & 27.7 & $7.56_{-0.33}^{+0.48}$ & $-19.32_{-0.03}^{+0.03}$ \\
beacon\_0447-2637-2129 & $71.703773$ & $-26.632273$ & 27.1 & 26.5 & $7.56_{-0.07}^{+0.08}$ & $-19.97_{-0.01}^{+0.01}$ \\
beacon\_0217-0509-1974 & $34.221172$ & $-5.180426$ & 27.4 & 27.8 & $7.57_{-0.22}^{+0.22}$ & $-19.77_{-0.01}^{+0.01}$ \\
beacon\_0217-0509-6250 & $34.214062$ & $-5.142040$ & 28.4 & 28.2 & $7.59_{-0.46}^{+0.53}$ & $-18.80_{-0.05}^{+0.04}$ \\
beacon\_0217-0509-12575 & $34.207989$ & $-5.121376$ & 28.4 & 28.0 & $7.65_{-0.37}^{+0.37}$ & $-18.81_{-0.03}^{+0.04}$ \\
beacon\_0217-0509-85 & $34.234711$ & $-5.188540$ & 28.0 & 27.6 & $7.65_{-0.46}^{+0.55}$ & $-19.28_{-0.06}^{+0.05}$ \\
beacon\_0217-0509-1420 & $34.225983$ & $-5.182612$ & 28.6 & 28.1 & $7.66_{-0.59}^{+0.70}$ & $-18.65_{-0.09}^{+0.08}$ \\
beacon\_1420+5252-12428 & $215.084641$ & $52.882202$ & 27.9 & 26.9 & $7.70_{-0.05}^{+0.05}$ & $-19.54_{-0.01}^{+0.01}$ \\
beacon\_0217-0508-3329 & $34.318546$ & $-5.148247$ & 28.4 & 28.8 & $7.71_{-0.44}^{+0.47}$ & $-18.73_{-0.04}^{+0.05}$ \\
beacon\_0217-0504-6193 & $34.273129$ & $-5.037531$ & 28.3 & 27.6 & $7.72_{-0.46}^{+0.50}$ & $-19.04_{-0.04}^{+0.05}$ \\
beacon\_0014-3025-6550 & $3.572950$ & $-30.413603$ & -- & 28.8 & $7.77_{-0.62}^{+0.63}$ & $-17.97_{-0.07}^{+0.06}$ \\
beacon\_0217-0508-8355 & $34.290524$ & $-5.122096$ & 28.8 & 28.8 & $7.81_{-0.54}^{+0.52}$ & $-18.25_{-0.05}^{+0.05}$ \\
beacon\_0332-2745-9353 & $53.022465$ & $-27.739815$ & 28.9 & 28.5 & $7.82_{-0.50}^{+0.46}$ & $-18.32_{-0.04}^{+0.04}$ \\
beacon\_0332-2745-15255 & $53.043163$ & $-27.734766$ & 29.0 & 27.8 & $7.83_{-0.77}^{+0.78}$ & $-18.30_{-0.09}^{+0.11}$ \\
beacon\_0217-0508-8334 & $34.286720$ & $-5.123929$ & 29.5 & 29.2 & $7.84_{-0.67}^{+0.70}$ & $-17.68_{-0.07}^{+0.06}$ \\
beacon\_0217-0508-14393 & $34.269173$ & $-5.105772$ & 28.1 & 26.8 & $7.88_{-0.44}^{+0.39}$ & $-19.06_{-0.08}^{+0.05}$ \\
beacon\_0014-3025-14457 & $3.608568$ & $-30.418518$ & -- & 25.8 & $7.90_{-0.48}^{+0.19}$ & $-21.17_{-0.02}^{+0.02}$ \\
beacon\_0332-2745-18704 & $53.035835$ & $-27.718058$ & 28.5 & 28.0 & $7.90_{-0.77}^{+0.69}$ & $-18.72_{-0.09}^{+0.07}$ \\
beacon\_0217-0509-9363 & $34.219463$ & $-5.127748$ & 27.9 & 26.9 & $7.90_{-0.52}^{+0.33}$ & $-19.36_{-0.06}^{+0.04}$ \\
beacon\_0217-0509-2346 & $34.244049$ & $-5.167047$ & 29.2 & 28.5 & $7.94_{-0.81}^{+0.81}$ & $-18.08_{-0.13}^{+0.09}$ \\
beacon\_0217-0504-15163 & $34.292065$ & $-5.043071$ & 27.8 & 27.8 & $7.96_{-0.38}^{+0.31}$ & $-19.39_{-0.01}^{+0.04}$ \\
beacon\_0217-0504-10020 & $34.314068$ & $-5.089706$ & 28.3 & 26.3 & $7.99_{-0.83}^{+0.74}$ & $-18.77_{-0.16}^{+0.12}$ \\
beacon\_0217-0508-3572 & $34.296047$ & $-5.156882$ & 28.1 & 27.9 & $8.03_{-0.58}^{+0.46}$ & $-19.12_{-0.04}^{+0.05}$ \\
beacon\_0217-0508-1995 & $34.301559$ & $-5.166206$ & 27.4 & 26.6 & $8.06_{-0.36}^{+0.34}$ & $-19.94_{-0.03}^{+0.03}$ \\
beacon\_0217-0504-3421 & $34.270123$ & $-5.040110$ & 27.6 & 26.9 & $8.06_{-0.78}^{+0.33}$ & $-19.57_{-0.05}^{+0.05}$ \\
beacon\_0014-3025-2599 & $3.553823$ & $-30.410107$ & -- & 28.1 & $8.07_{-0.57}^{+0.37}$ & $-18.68_{-0.03}^{+0.04}$ \\
beacon\_0959+0200-8926 & $149.829056$ & $2.021058$ & 27.9 & 26.5 & $8.10_{-0.21}^{+0.35}$ & $-19.38_{-0.02}^{+0.03}$ \\
beacon\_1420+5252-11933 & $215.035583$ & $52.892235$ & 25.2 & 24.7 & $8.12_{-0.12}^{+0.11}$ & $-22.05_{-0.01}^{+0.01}$ \\
beacon\_0217-0509-1475 & $34.246601$ & $-5.172431$ & 28.0 & 27.9 & $8.12_{-0.83}^{+0.41}$ & $-19.19_{-0.06}^{+0.05}$ \\
beacon\_0014-3025-8648 & $3.613551$ & $-30.434706$ & -- & 27.2 & $8.18_{-0.25}^{+0.25}$ & $-19.70_{-0.01}^{+0.01}$ \\
beacon\_1010+2701-2444 & $152.446686$ & $26.987419$ & 27.2 & 26.6 & $8.23_{-1.00}^{+0.66}$ & $-20.08_{-0.13}^{+0.06}$ \\
beacon\_0217-0509-9252 & $34.209660$ & $-5.132755$ & 29.4 & 28.5 & $8.24_{-0.93}^{+0.75}$ & $-17.82_{-0.09}^{+0.07}$ \\
beacon\_1010+2701-19789 & $152.471619$ & $27.051905$ & 28.0 & 27.6 & $8.26_{-0.93}^{+0.61}$ & $-19.22_{-0.05}^{+0.05}$ \\
beacon\_0332-2745-17578 & $53.037659$ & $-27.711735$ & 27.3 & 26.5 & $8.27_{-1.02}^{+0.29}$ & $-20.08_{-0.21}^{+0.01}$ \\
beacon\_0332-2749-18718 & $53.058949$ & $-27.808054$ & 27.1 & 26.3 & $8.27_{-0.19}^{+0.17}$ & $-20.29_{-0.01}^{+0.01}$ \\
beacon\_1010+2701-16958 & $152.464661$ & $27.046890$ & 27.8 & 27.5 & $8.28_{-1.04}^{+0.61}$ & $-19.40_{-0.12}^{+0.05}$ \\
beacon\_0332-2749-24726 & $53.001209$ & $-27.802843$ & 26.8 & 25.3 & $8.29_{-0.13}^{+0.12}$ & $-20.74_{-0.01}^{+0.02}$ \\
beacon\_0217-0504-3873 & $34.314297$ & $-5.105042$ & 28.9 & 28.5 & $8.30_{-0.86}^{+0.51}$ & $-18.34_{-0.06}^{+0.05}$ \\
beacon\_0332-2745-8569 & $53.053482$ & $-27.765987$ & 28.3 & 27.8 & $8.35_{-0.40}^{+0.35}$ & $-18.83_{-0.02}^{+0.02}$ \\
beacon\_0217-0509-1793 & $34.224037$ & $-5.180541$ & 27.9 & 26.1 & $8.39_{-0.22}^{+0.26}$ & $-19.50_{-0.01}^{+0.01}$ \\
beacon\_0332-2745-8259 & $53.050213$ & $-27.764677$ & 27.0 & 26.3 & $8.40_{-0.35}^{+0.54}$ & $-20.29_{-0.05}^{+0.06}$ \\
beacon\_0217-0509-336 & $34.223042$ & $-5.191769$ & 26.0 & 24.9 & $8.41_{-0.13}^{+0.13}$ & $-21.52_{-0.01}^{+0.01}$ \\
beacon\_0217-0508-1427 & $34.307549$ & $-5.168149$ & 27.9 & 27.5 & $8.41_{-0.32}^{+0.26}$ & $-19.40_{-0.01}^{+0.03}$ \\
beacon\_0332-2745-17579 & $53.037807$ & $-27.711599$ & 26.7 & 26.5 & $8.41_{-0.22}^{+0.21}$ & $-20.63_{-0.01}^{+0.01}$ \\
beacon\_0332-2745-9828 & $53.012547$ & $-27.731163$ & 27.0 & 26.1 & $8.42_{-0.14}^{+0.14}$ & $-20.34_{-0.01}^{+0.01}$ \\
beacon\_0217-0508-9685 & $34.272381$ & $-5.125317$ & 28.9 & 28.8 & $8.45_{-0.57}^{+0.38}$ & $-18.46_{-0.04}^{+0.03}$ \\
beacon\_0014-3025-11608 & $3.612495$ & $-30.427191$ & -- & 27.2 & $8.47_{-0.26}^{+0.22}$ & $-19.39_{-0.01}^{+0.01}$ \\
beacon\_0217-0509-435 & $34.222919$ & $-5.191683$ & 28.3 & 27.2 & $8.49_{-1.15}^{+0.30}$ & $-19.19_{-0.32}^{+0.01}$ \\
beacon\_0217-0509-6946 & $34.211567$ & $-5.140682$ & 28.2 & 27.4 & $8.52_{-0.45}^{+0.38}$ & $-19.12_{-0.01}^{+0.05}$ \\
beacon\_0217-0504-24295 & $34.306004$ & $-5.045831$ & 28.9 & 28.0 & $8.52_{-0.75}^{+0.65}$ & $-18.43_{-0.07}^{+0.06}$ \\
beacon\_0217-0508-3135 & $34.317574$ & $-5.149925$ & 28.5 & 27.8 & $8.54_{-0.40}^{+0.33}$ & $-18.88_{-0.02}^{+0.02}$ \\
beacon\_0217-0509-14993 & $34.214493$ & $-5.112545$ & 27.4 & 27.4 & $8.56_{-0.30}^{+0.23}$ & $-19.88_{-0.01}^{+0.01}$ \\
beacon\_0217-0509-13335 & $34.218163$ & $-5.103996$ & 28.3 & 27.9 & $8.59_{-0.40}^{+0.30}$ & $-19.13_{-0.01}^{+0.02}$ \\
beacon\_0217-0504-3515 & $34.270214$ & $-5.040026$ & 27.7 & 26.7 & $8.63_{-0.19}^{+0.17}$ & $-19.79_{-0.01}^{+0.01}$ \\
beacon\_0332-2749-12382 & $53.087475$ & $-27.814884$ & 27.3 & 26.0 & $8.64_{-0.16}^{+0.12}$ & $-20.18_{-0.01}^{+0.01}$ \\
beacon\_0217-0508-9763 & $34.272213$ & $-5.124932$ & 26.6 & 26.1 & $8.68_{-0.12}^{+0.14}$ & $-20.86_{-0.01}^{+0.01}$ \\
beacon\_0447-2637-11548 & $71.671661$ & $-26.582500$ & 27.3 & 27.6 & $8.73_{-0.29}^{+0.33}$ & $-19.97_{-0.03}^{+0.01}$ \\
beacon\_0217-0508-8040 & $34.302929$ & $-5.117467$ & 28.1 & 27.8 & $8.75_{-0.30}^{+0.28}$ & $-19.29_{-0.01}^{+0.02}$ \\
beacon\_0217-0508-8415 & $34.292923$ & $-5.120759$ & 28.4 & 28.6 & $8.77_{-0.40}^{+0.31}$ & $-18.84_{-0.01}^{+0.03}$ \\
beacon\_0860+3857-14145 & $134.982574$ & $38.985649$ & 26.3 & 25.7 & $8.79_{-0.21}^{+0.19}$ & $-21.08_{-0.01}^{+0.01}$ \\
beacon\_0014-3025-10503 & $3.624762$ & $-30.437508$ & -- & 27.0 & $8.81_{-0.17}^{+0.16}$ & $-20.22_{-0.01}^{+0.01}$ \\
beacon\_0217-0509-4602 & $34.220940$ & $-5.160698$ & 27.3 & 26.2 & $8.82_{-0.16}^{+0.25}$ & $-20.12_{-0.01}^{+0.02}$ \\
beacon\_1010+2701-420 & $152.422150$ & $26.986328$ & 26.0 & 24.7 & $8.87_{-0.25}^{+0.31}$ & $-21.44_{-0.03}^{+0.04}$ \\
beacon\_0217-0508-12114 & $34.277187$ & $-5.113205$ & 28.3 & 27.0 & $8.92_{-0.23}^{+0.25}$ & $-19.21_{-0.01}^{+0.01}$ \\
beacon\_0217-0509-11648 & $34.212341$ & $-5.122705$ & 27.7 & 27.2 & $8.98_{-0.20}^{+0.21}$ & $-19.57_{-0.01}^{+0.01}$ \\
beacon\_1420+5252-2768 & $215.008652$ & $52.868328$ & 26.4 & 26.4 & $9.04_{-0.15}^{+0.14}$ & $-21.05_{-0.01}^{+0.01}$ \\
beacon\_0217-0504-10016 & $34.283627$ & $-5.044103$ & 28.0 & 27.9 & $9.07_{-0.22}^{+0.24}$ & $-19.42_{-0.01}^{+0.01}$ \\
beacon\_0959+0200-16783 & $149.836929$ & $2.041479$ & 26.9 & 27.2 & $9.14_{-0.21}^{+0.17}$ & $-20.52_{-0.01}^{+0.01}$ \\
beacon\_0217-0504-7157 & $34.309956$ & $-5.090462$ & 26.9 & 27.7 & $9.16_{-0.15}^{+0.15}$ & $-20.28_{-0.02}^{+0.01}$ \\
beacon\_0217-0509-14115 & $34.200089$ & $-5.116339$ & 28.7 & 29.7 & $9.25_{-0.38}^{+0.09}$ & $-18.55_{-0.01}^{+0.01}$ \\
beacon\_1010+2701-18712 & $152.439987$ & $27.060465$ & 26.9 & 25.9 & $9.31_{-0.42}^{+1.60}$ & $-20.63_{-0.10}^{+0.24}$ \\
beacon\_0217-0508-1490 & $34.314037$ & $-5.164579$ & 28.3 & 28.1 & $9.38_{-0.55}^{+0.34}$ & $-19.25_{-0.06}^{+0.01}$ \\
beacon\_0217-0508-3296 & $34.305820$ & $-5.154357$ & 27.3 & 27.5 & $9.40_{-0.16}^{+0.28}$ & $-20.14_{-0.01}^{+0.01}$ \\
beacon\_0014-3025-13872 & $3.617229$ & $-30.425524$ & -- & 26.0 & $9.48_{-0.08}^{+0.08}$ & $-21.69_{-0.01}^{+0.01}$ \\
beacon\_0217-0504-9016 & $34.283463$ & $-5.046202$ & 28.2 & 28.0 & $9.52_{-1.57}^{+1.32}$ & $-19.19_{-0.52}^{+0.15}$ \\
beacon\_0217-0509-2696 & $34.237087$ & $-5.167516$ & 27.6 & 27.2 & $9.52_{-1.46}^{+1.34}$ & $-19.82_{-0.54}^{+0.15}$ \\
beacon\_1010+2701-17114 & $152.439484$ & $27.056280$ & 26.5 & 26.0 & $9.67_{-0.22}^{+1.26}$ & $-21.42_{-0.18}^{+0.07}$ \\
beacon\_0332-2749-16743 & $53.001728$ & $-27.812489$ & 28.1 & 27.7 & $9.79_{-0.26}^{+0.57}$ & $-19.75_{-0.01}^{+0.04}$ \\
beacon\_0442-4613-8264 & $70.427460$ & $-46.230682$ & 26.2 & 26.9 & $9.83_{-0.99}^{+0.69}$ & $-21.02_{-0.07}^{+0.04}$ \\
beacon\_0217-0508-7634 & $34.305424$ & $-5.118019$ & 28.5 & 28.7 & $9.91_{-0.72}^{+0.78}$ & $-18.92_{-0.06}^{+0.05}$ \\
beacon\_0860+3857-10356 & $134.977585$ & $38.971680$ & 27.5 & 26.5 & $9.99_{-0.97}^{+1.22}$ & $-19.98_{-0.18}^{+0.14}$ \\
beacon\_0217-0509-12959 & $34.221210$ & $-5.113707$ & 29.0 & 28.6 & $10.02_{-1.03}^{+1.11}$ & $-18.62_{-0.13}^{+0.08}$ \\
beacon\_0217-0508-11070 & $34.298710$ & $-5.107352$ & 27.8 & 27.8 & $10.04_{-1.03}^{+0.61}$ & $-19.66_{-0.07}^{+0.06}$ \\
beacon\_0332-2749-19513 & $53.075996$ & $-27.806505$ & 27.8 & 28.0 & $10.13_{-0.43}^{+0.42}$ & $-19.74_{-0.02}^{+0.01}$ \\
beacon\_1138+5748-2321 & $174.362198$ & $57.776512$ & 27.1 & 26.3 & $10.17_{-1.31}^{+1.36}$ & $-20.52_{-0.15}^{+0.15}$ \\
beacon\_0332-2749-17529 & $53.085506$ & $-27.808601$ & 27.7 & 28.0 & $10.24_{-0.62}^{+0.67}$ & $-19.90_{-0.04}^{+0.03}$ \\
beacon\_0217-0509-13052 & $34.221394$ & $-5.113218$ & 28.1 & 28.4 & $10.26_{-0.61}^{+0.45}$ & $-19.49_{-0.02}^{+0.04}$ \\
beacon\_0217-0509-11058 & $34.217091$ & $-5.122624$ & 28.7 & 27.3 & $10.32_{-1.11}^{+1.12}$ & $-18.84_{-0.17}^{+0.13}$ \\
beacon\_1010+2701-12141 & $152.432724$ & $27.044855$ & 28.4 & 26.7 & $10.75_{-1.30}^{+0.98}$ & $-19.34_{-0.08}^{+0.12}$ \\
beacon\_0959+0200-14138 & $149.845947$ & $2.032689$ & 28.2 & 27.5 & $11.04_{-0.38}^{+0.44}$ & $-19.57_{-0.01}^{+0.03}$ \\
beacon\_0332-2745-18596 & $53.085110$ & $-27.757654$ & 28.2 & 27.8 & $11.22_{-0.70}^{+0.70}$ & $-20.07_{-0.04}^{+0.04}$ \\
beacon\_0332-2749-6608 & $53.005013$ & $-27.824554$ & 29.3 & 27.8 & $12.29_{-1.19}^{+2.46}$ & $-19.64_{-0.14}^{+0.24}$ \\
\enddata
\tablecomments{Sources are sorted in increased order of redshift.}
\label{tab:src}
\end{deluxetable*}

\begin{deluxetable*}{lcl}
\tabletypesize{\footnotesize}
\tabcolsep=6pt
\tablecolumns{3}
\tablewidth{0pt} 
\tablecaption{Columns and description for the photometric catalogs.}
\tablehead{
\colhead{Column} & \colhead{Unit} & \colhead{Description}
}
\startdata
\cutinhead{Photometric Catalog}
id & & Individual identification name of objects.\\
ra &degree & R.A. (J2000).\\
dec &degree & Declination (J2000).\\
x &pixel & x position in image coordinate.\\
y &pixel & y position in image coordinate.\\
f\_{\ttfamily xx} &$f_{\nu}$  & Total flux of objects in band {\ttfamily xx}.\\
e\_{\ttfamily xx} &$f_{\nu}$  & Flux error ($1\sigma$) in total flux of objects in band {\ttfamily xx}.\\
flux\_iso\_{\ttfamily xx} &$f_{\nu}$  & Isophotal flux in band {\ttfamily xx}.\\
fluxerr\_iso\_{\ttfamily xx} &$f_{\nu}$  & Flux error ($1\sigma$) in isophotal flux in band {\ttfamily xx}.\\
flux\_aper\_{\ttfamily n}\_{\ttfamily xx} &$f_{\nu}$  & Aperture flux in band {\ttfamily xx} in the {\ttfamily n}\,th aperture. Aperture diameter sizes are: [0.16, 0.32, 0.48, 0.64, 1.28, 2.56]\,arcsec.\\
flux\_aper\_{\ttfamily n}\_{\ttfamily xx} &$f_{\nu}$  & Flux error ($1\sigma$) in aperture flux in band {\ttfamily xx} in the {\ttfamily n}\,th aperture.\\
flag\_{\ttfamily xx} & & SExtractor photometry flag for band {\ttfamily xx}.\\
kron\_radius\_{\ttfamily xx} &pixel & Kron radius in band {\ttfamily xx}.\\
a\_image\_{\ttfamily xx} &pixel & Radius along the major axis in band {\ttfamily xx}.\\
b\_image\_{\ttfamily xx} &pixel & Radius along the minor axis in band {\ttfamily xx}.\\
theta\_{\ttfamily xx} &degree & SExtractor position angle in band {\ttfamily xx} measured counterclockwise.\\
class\_star\_{\ttfamily xx} & & SExtractor class star indicator in band {\ttfamily xx}.\\
flux\_radius\_{\ttfamily xx} &pixel & SExtractor Non-parametric half-light radius in band {\ttfamily xx}.\\
flux\_scale & & Scale ratio of total flux to aperture flux measured in the detection band. The minimum scale is set to 1.\\
\cutinhead{Redshift catalog}
z{\ttfamily n} & & Photometric redshift at the {\ttfamily n}\,th percentile of redshift likelihood distribution.\\
zpeak & & Photometric redshift at the peak of global likelihood redshift distribution.\\
chi2 & & Reduced chi-square at zpeak.\\
MUV{\ttfamily n} &mag & Absolute UV (1450\,\AA) magnitude calculated at z{\ttfamily n}.\\
MUVpeak &mag & Absolute UV (1450\,\AA) magnitude calculated at zpeak.\\
plow & & Total probability of redshift distribution at $z<{\rm zset}$.\\
phigh & & Total probability of redshift distribution at $z>{\rm zset}$.\\
UVbeta\_lambda & & UV-beta slope ($\beta_\lambda$) calculated with the best-fit template at zpeak.\\
U-V &mag & Rest-frame $U-V$ color calculated with the best-fit template at zpeak.\\
B-V &mag & Rest-frame $B-V$ color calculated with the best-fit template at zpeak.\\
V-J &mag & Rest-frame $V-J$ color calculated with the best-fit template at zpeak.\\
z-J &mag & Rest-frame $z-J$ color calculated with the best-fit template at zpeak.\\
z\_peak\_low & & Photometric redshift at the peak of redshift likelihood distribution at $z<z_{\rm set}$.\\
chi2\_peak\_low & & Reduced chi-square at z\_peak\_low.\\
z\_peak\_high & & Photometric redshift at the peak of redshift likelihood distribution at $z>z_{\rm set}$.\\
chi2\_peak\_high & & Reduced chi-square at z\_peak\_high.\\
nfilt & & Number of filters used for photometric redshift fit.
\enddata
\tablecomments{All fluxes are corrected for the Galactic dust extinction, and set to magnitude zeropoint of $m_0=25$ i.e. $m=-2.5 \log(f_\nu)+m_0$. For sources with filters without data coverage, the flux error arrays are set to $-99$.
}
\label{tab:col}
\end{deluxetable*}



\bibliography{ms}{}
\bibliographystyle{aasjournal}



\end{document}